\documentclass[a4paper,11pt]{article}

\usepackage{jheppub}

\usepackage{hyperref}
\usepackage{graphicx}
\usepackage{amsmath}
\usepackage{amssymb}
\usepackage{xspace}
\usepackage[small]{subfigure}
\usepackage{slashed}
\usepackage[normalem]{ulem}
\usepackage{xcolor} 

\newcommand{\eq}[1]{Eq.~\eqref{eq:#1}}
\newcommand{\eqs}[2]{Eqs.~\eqref{eq:#1} and \eqref{eq:#2}}
\renewcommand{\sec}[1]{Sec.~\ref{sec:#1}}

\newcommand{\app}[1]{App.~\ref{app:#1}}
\newcommand{\fig}[1]{Fig.~\ref{fig:#1}}

\usepackage{marginnote}

\newcommand{\df}{\mathrm{d}}

\newcommand{\plus}{\!+\!}
\newcommand{\minus}{\!-\!}

\newcommand{\Jht}{{J_{B}}}

\newcommand{\bn}{{\bar{n}}}

\newcommand{\GeV}{\,\mathrm{GeV}}

\newcommand{\nn}{\nonumber}

\newcommand{\bare}{\mathrm{bare}}

\newcommand{\SCETa}{\ensuremath{{\rm SCET}_{\rm I}}\xspace}
\newcommand{\SCETb}{\ensuremath{{\rm SCET}_{\rm II}}\xspace}

\def\lmu{L_{m}}
\def\lmubar{L_{\bar m}}
\def\lqp{L_Q}
\def\ln{\textrm{ln}}
\def\df{\textrm{d}}
\def\nn{\nonumber}
\def\MS{\overline{\rm MS}}

\def\LQCD{\Lambda_{\rm QCD}}

\def\bare{{\textrm{bare}}}
\def\SCET{{\rm{SCET}}}

\allowdisplaybreaks[2]

\preprint{\begin{flushright}
DESY  15-139 \\
MIT--CTP 4693\\ 
UWThPh-2015-18
\end{flushright}}

\title{Hard Matching for Boosted Tops at Two Loops}

\author[a,b]{Andr\'e H.~Hoang,}
\affiliation[a]{University of Vienna, Faculty of Physics, Boltzmanngasse 5, A-1090 Wien, Austria}
\affiliation[b]{Erwin Schr\"odinger International Institute for Mathematical Physics, University of Vienna, Boltzmanngasse 9, A-1090 Vienna, Austria}

\author[c]{Aditya Pathak,}
\affiliation[c]{Center for Theoretical Physics, Massachusetts Institute of Technology, Cambridge, MA~02139, U.S.A.}

\author[d]{Piotr Pietrulewicz,}
\affiliation[d]{Theory Group, Deutsches Elektronen-Synchrotron (DESY), D-22607 Hamburg, Germany}

\author[c]{Iain W.~Stewart}

\emailAdd{andre.hoang@univie.ac.at}
\emailAdd{adityap@mit.edu}
\emailAdd{piotr.pietrulewicz@desy.de}
\emailAdd{iains@mit.edu}

\abstract{
Cross sections for top quarks provide very interesting physics opportunities, being both sensitive to new physics and also perturbatively tractable due to the large top quark mass. Rigorous factorization theorems for top cross sections can be derived in several kinematic scenarios, including the boosted regime in the peak region that we consider here. In the context of the corresponding factorization theorem for $e^+e^-$ collisions we extract the last missing ingredient that is needed to evaluate the cross section differential in the jet-mass at two-loop order, namely the matching coefficient at the scale $\mu\simeq m_t$. Our extraction also yields the final ingredients needed to carry out logarithmic resummation at  next-to-next-to-leading logarithmic order (or N$^3$LL if we ignore the missing 4-loop cusp anomalous dimension).
This coefficient exhibits an amplitude level rapidity logarithm starting at $\mathcal{O}(\alpha_s^2)$ due to virtual top quark loops, which we treat using rapidity renormalization group (RG) evolution. Interestingly, this rapidity RG evolution appears in the matching coefficient between two effective theories around the heavy quark mass scale $\mu\simeq m_t$.
}


\begin{document}
\maketitle

\pagebreak
\section{Introduction}
\label{sec:Intro}

The top quark mass is one of the most important parameters in the Standard Model. As the heaviest observed fermion, the top quark provides an important probe for the Higgs sector, and gives dominant contributions to many electroweak observables, thus providing strong benchmark constraints for extensions of the Standard Model. Furthermore, the mass of the top quark and the Higgs boson represent crucial parameters in studies of the stability of the Standard Model vacuum~\cite{Cabibbo:1979ay,Alekhin:2012py,Buttazzo:2013uya,Andreassen:2014gha,Branchina:2013jra,Branchina:2014usa}. Precision measurements of the top quark mass are a difficult task due to challenges from both experimental and theoretical sides, mainly related to the fact that the top quark is a colored particle. 

The current value of the top quark mass from a combined analysis of Tevatron and LHC data is $m_t=173.34 \pm 0.76$ GeV~\cite{ATLAS:2014wva}, see also~\cite{CMS:2014hta,Aad:2015nba}. The precision obtained in this result relies on Monte Carlo (MC) based template and matrix element methods, which aim  to account for essentially all of the kinematic final state information in the top quark events. However, this approach does not account for the relation of  the extracted MC top quark parameter to an unambiguous field theoretic QCD top mass definition~\cite{Hoang:2008xm,Moch:2014tta,Hoang:2014oea}. At the time of  writing, no procedure to systematically quantify and improve this relation exists. While it seems unlikely that the template and matrix element analyses can be based on first principle QCD calculations which can be systematically improved to specify the top mass scheme unambiguously, it is quite plausible that other highly sensitive top mass observables can be devised which can clarify the issue by making high precision theoretical calculations feasible. 

One method to determine $m_t$ in a well-defined mass scheme from a kinematic spectrum with small uncertainties has been discussed in Refs.~\cite{Fleming:2007qr,Fleming:2007xt,Hoang:2008xm}. Here the hemisphere dijet invariant mass distribution in the peak region for the production of boosted tops in electron-positron annihilation was suggested as an observable and it was shown that hadron level predictions of the double differential distribution can be carried out in a stable manner within a constrained set of top quark mass schemes. It was in particular demonstrated that the location of the peak of the distribution is highly sensitive to the top quark mass, and that only specific low-scale short-distance mass definitions are suitable for high-precision extractions. Although the effective theory setup developed therein was devised for the context of a future $e^+e^-$ collider, the approach can be extended to the environment at hadron colliders taking into account the complications related to initial state radiation, underlying event, parton distribution functions and dependence on jet algorithms and jet radius~\cite{Hoang:2015pptop}. In Refs.~\cite{Fleming:2007qr,Fleming:2007xt} the calculation for $e^+e^-$ annihilation was carried out at Next-to-Leading Logarithmic (NLL) accuracy with the perturbative ingredients at ${\cal{O}}(\alpha_s)$. In this paper we provide a result for the ${\cal O}(\alpha_s^2)$ matching correction at the scale $\mu\simeq m_t$ for the $e^+e^-$-collider setup. Taken together with the known ${\cal O}(\alpha_s^2)$ results for the jet function in the heavy-quark limit from Ref.~\cite{Jain:2008gb}, for the massless soft function from Refs.~\cite{Kelley:2011ng,Hornig:2011iu,Monni:2011gb}, and input from previous form factor calculations for massless quark production~\cite{Matsuura:1987wt,Matsuura:1988sm}, our result provides the last missing ingredient needed to extend the $e^+e^-$ boosted top jet analysis to ${\cal{O}}(\alpha_s^2)$. In turn, with known results, these fixed order contributions can be accompanied with resummation of logarithms up to next-to-next-to-leading logarithmic order (NNLL). Up to the missing four loop cusp anomalous dimension, which is known to give a very small correction (see e.g.~\cite{Becher:2008cf,Abbate:2010xh}), all ingredients are also available for N$^3$LL. 

Boosted top quark production with subsequent decays in the peak region of the invariant mass distributions involves physical effects in a range of widely separated energy scales. The hierarchy between the production energy $Q$, the top mass $m_t$, the decay width $\Gamma_t$ and the hadronization scale $\LQCD$ is given by $Q \gg m_t \gg \Gamma_t > \LQCD$. Given this hierarchy of scales, the cross section contains large logarithms of ratios of these scales which spoil the perturbative expansion in $\alpha_s$. This necessitates to replace fixed order computations by resummed calculations. The Effective Field Theory (EFT) setup devised in Ref.~\cite{Fleming:2007qr,Fleming:2007xt} disentangles the fluctuations at the different scales and allows us to resum the logarithms through renormalization group evolution (RGE).\footnote{Note that our boosted top limit differs from the application of HQET in Ref.~\cite{vonManteuffel:2014mva}, which considers $t\bar t$ production with slow top quarks. It also differs from the work of Ref.~\cite{Ferroglia:2012uy}, which considers the top-pair invariant mass for boosted top quarks, rather than the individual boosted top jets. Hence the factorization theorem for our case differs from the ones considered there.}

\begin{figure}
 \begin{minipage}{6cm}
 \centering
\includegraphics[height=1.3\linewidth]{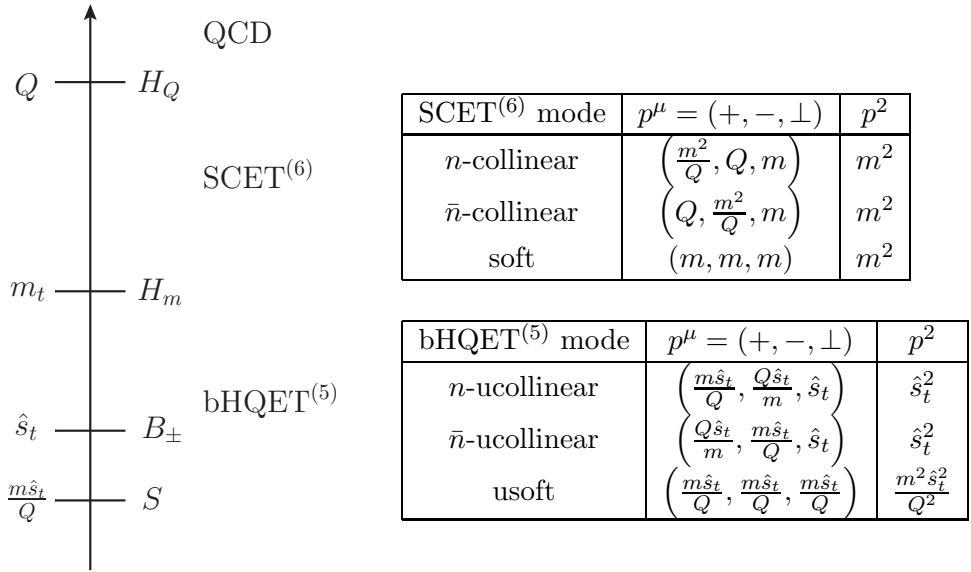}%
\end{minipage}
 \begin{minipage}{10cm}
 \vspace{0.5cm}
\begin{tabular}{|c|c|c|c|}
\hline
SCET$^{(6)}$ mode & $p^\mu=(+,-,\perp)$ & $p^2$ \\ 
\hline
$n$-collinear &  $\Big(\frac{m^2}{Q},Q,m\Big)$ &$m^2$ \\ 
$\bar{n}$-collinear &  $\Big(Q,\frac{m^2}{Q},m\Big)$ &$m^2$\\
soft & $ \left(m,m,m\right)$ &$m^2$ \\ 
\hline
\end{tabular}  
\vspace{0.5cm}
\newline
\begin{tabular}{|c|c|c|c|}
\hline
bHQET$^{(5)}$ mode  & $p^\mu=(+,-,\perp)$ & $p^2$ \\ 
\hline
$n$-ucollinear &  $\Big(\frac{m \hat{s}_t}{Q},\frac{Q \hat{s}_t}{m},\hat{s}_t\Big)$  &$\hat{s}^2_t$ \\ 
$\bar{n}$-ucollinear &  $\Big(\frac{Q \hat{s}_t}{m},\frac{m \hat{s}_t}{Q},\hat{s}_t\Big)$ &$\hat{s}^2_t$\\
usoft & $ \left(\frac{m \hat{s}_t}{Q},\frac{m \hat{s}_t}{Q},\frac{m \hat{s}_t}{Q}\right)$ &$\frac{m^2 \hat{s}_t^2}{Q^2}$ \\ 
\hline
\end{tabular}
\end{minipage}
   \caption{Scales and effective theories with associated structures in the factorization theorem for boosted top production ($Q\gg m_t$) with jet invariant masses close to the top mass. The superscripts $(5)$ and $(6)$ indicate the number of dynamic flavors in the theory. Note that in this context SCET just plays a role of an intermediate EFT with all invariant mass fluctuations above or of order the mass scale, in which the observable is not yet measured. For definiteness we also display the scaling of the EFT modes in light-cone coordinates.} 
  \label{fig:scales_bHQET}
\end{figure}

We are interested in the peak region where each of the jet invariant masses, for the top $s_t$ and antitop $s_{\bar{t}}$, is close to the top quark mass, i.e., 
\begin{align}
 \hat{s}_{t,\bar{t}} \equiv \frac{s_{t,\bar{t}} - m_t^2}{m_t} \ll m_t 
 \,.
\end{align}
For this kinematic region both of the hierarchies $\hat{s}_{t,\bar{t}} \sim \Gamma_t$ and $\hat{s}_{t,\bar{t}}  \gg \Gamma_t$ are allowed.  The sequence of the EFTs and the corresponding modes relevant for this problem are displayed in Fig.~\ref{fig:scales_bHQET}. First, hard modes with fluctuations with virtualities of order $\sim Q$ are integrated out in QCD. The corresponding low-energy theory containing collinear and soft modes is Soft Collinear Effective Theory (SCET)~\cite{Bauer:2000ew,Bauer:2000yr,Bauer:2001ct,Bauer:2001yt}, which allows to resum large logarithms between $Q$ and $m_t$. In a second step all fluctuations with virtualities of order $\sim m_t$ are integrated out, and SCET is thus matched onto boosted Heavy Quark Effective Theory (bHQET), an EFT with ultracollinear and ultrasoft modes at a lower invariant mass scale, which allows to resum logarithms between $m_t$ and $\hat{s}_{t,\bar{t}}$. The factorization theorem for the double differential cross section in $e^+ e^-$ collisions reads
\begin{align}
\label{eq:dijetobservable}
\frac{1}{\sigma_0}\frac{d \sigma}{ \df s_t \, \df s_{\bar{t}}} = & \,  H_Q\left(Q,\mu\right) H_m \bigg (m_t, \frac{Q}{m_t},\mu \bigg )
 \int d \ell^+ d \ell^- S(\ell^+,\ell^-,\mu) 
  \nn\\
 & \times \Jht \bigg (\frac{{s}_t -m_t^2 - Q \ell^+}{m_t}, \Gamma_t, \delta m,\mu \bigg ) \Jht \bigg ( \frac{{s}_{\bar{t}}-m_t^2 - Q \ell^-}{m_t}, \Gamma_t, \delta m, \mu \bigg) 
 \nn \\
& \times \bigg[ 1 +{\cal{O}} \bigg ( \frac{m_t\alpha_s}{Q} \bigg ) + {\cal{O}} \bigg ( \frac{m_t^2}{Q^2} \bigg) + {\cal{O}} \bigg ( \frac{\Gamma_t}{m_t} \bigg ) + {\cal{O}} \bigg( \frac{\hat{s}^2_{t,\bar{t}}}{m_t^2} \bigg ) \bigg] \, .
\end{align}
Here $\sigma_0$ denotes the tree level cross section for $e^+ e^- \to q\bar{q}$. The terms $H_Q$ and $H_m$ are hard functions related to the matching from QCD to SCET at the scale $\mu \sim Q$ and from SCET to bHQET at the scale $\mu \sim m$, respectively. The terms $\Jht$ and $S$ denote the jet and soft functions, respectively, which are nonlocal matrix elements in bHQET. Note that we use $\Jht$ for the heavy-quark jet function, rather than the symbol $B$ employed in Refs.~\cite{Fleming:2007qr,Fleming:2007xt,Jain:2008gb}. Here $\Jht$ describes the dynamics of the ultracollinear radiation inside the $t$ or $\bar{t}$ jet at the scale $\mu \sim \hat{s}_t$. The function $S$ incorporates the ultrasoft cross talk between the two jets at the scale $\mu \sim m \hat{s}_t/Q$, which is ${\cal O}(\LQCD)$ in the peak region, and perturbative in the tail above the peak. In Eq.~(\ref{eq:dijetobservable}) the RGE between the characteristic scale of each function and the common renormalization scale $\mu$ are implicit. We stress that in SCET the top quark is considered as dynamical and hence the RGE takes place with six active flavors, while for the ingredients that arise in bHQET there are only five dynamical flavors in the evolution. Note that it is possible that the ${\cal O}(m_t\alpha_s/Q)$ power corrections indicated in \eq{dijetobservable} are absent, but we are not aware of a rigorous proof at this time.

It is through the residual mass term $\delta m$ appearing in the bHQET jet functions $\Jht$ that the top quark mass scheme is specified unambiguously beyond tree-level. For order-by-order stable perturbative behavior, the top quark mass scheme employed should be free of the ${\cal O}(\Lambda_{\rm QCD})$ renormalon ambiguity, thus excluding the pole mass (specified by $\delta m =0$) as a choice. Furthermore, the parametric scaling of higher order corrections defining the mass scheme must be set by scales associated to the measurement, namely $\hat{s}_{t,\bar{t}},\Gamma_t\ll m_t$, in order not to violate the power counting required for the factorization.  This excludes employing the $\MS$ mass where these corrections scale as $\delta m \sim \alpha_s m_t$.  Valid options include the jet mass scheme~\cite{Fleming:2007qr,Fleming:2007xt,Jain:2008gb} or the MSR mass scheme~\cite{Hoang:2008xm,Jain:2008gb} which matches continuously onto $\overline{\rm MS}$. These two mass schemes have an adjustable cutoff parameter $R$ which controls the scaling of higher order corrections. 

The exact algorithm to determine the two jet regions and the precise form of the observable is irrelevant for the structure of Eq.~(\ref{eq:dijetobservable}) as long as parametrically $s_t\sim s_{\bar t}$, but does matter for the explicit perturbative expressions of its ingredients. The restriction $s_t\sim s_{\bar t}$ avoids large logarithms of the form $\ln(s_t/s_{\bar t})$, and is satisfied by variables designed to study the peak region of both jets, such as thrust.
In the analysis of Ref.~\cite{Fleming:2007xt} all particles were assigned to either of the two top jets depending on which hemisphere with respect to the thrust axis they enter. Thus the observable considered was physically close to event-shape distributions. The analysis of Ref.~\cite{Fleming:2007xt} for this inclusive jet observable was carried out at NLL$^\prime$, i.e. including perturbative ingredients at ${\cal{O}} (\alpha_s)$ and NLL resummation. At the time of writing the hard function $H_Q$, the bHQET jet function and the soft function are already known up to ${\cal{O}}(\alpha_s^2)$~\cite{Matsuura:1987wt,Jain:2008gb,Kelley:2011ng} or beyond, while resummation can be carried out to N$^3$LL.\footnote{So far the only missing ingredient for  N$^3$LL resummation (besides the four-loop cusp anomalous dimension) was the anomalous dimension of the  jet function $J_B$, or equivalently of the bHQET current, which we have now extracted from a recent result in literature in appendix~B.} Thus, the only relevant correction missing to perform a N$^3$LL analysis for the double hemisphere invariant mass distribution and similar observables in the peak region is the hard function $H_m$ at ${\cal{O}}(\alpha_s^2)$. This correction will affect the normalization of the differential cross section, while the shape of the cross section is determined mainly by the jet and soft functions. Here NNLL$^\prime$ refers to NNLL resummation with ${\cal O}(\alpha_s^2)$ fixed-order matching and matrix element corrections.

In this paper we carry out the computation of the ${\cal{O}}(\alpha_s^2)$ correction to $H_m$. In~\sec{HmSetup} we outline two methods to perform the computation. Instead of directly calculating the current matching factor between bHQET and SCET, we can also exploit the knowledge of the QCD heavy quark form factor calculated in Refs.~\cite{Bernreuther:2004ih,Gluza:2009yy} and various properties of the EFT to extract the hard function. In Sec.~\ref{sec:Hm} we carry out the computation at $\mathcal{O}(\alpha_s^2)$ using this method and show how to handle issues associated with the number of active quark flavors. This yields the result given in Eq.~(\ref{eq:Hm5f}) in terms of the pole mass.   In the two loop expression for $H_m$ we find terms of the form
\begin{align}\label{eq:rapidity_log}
 \alpha_s^2 C_F T_F \,\ln \bigg ( \frac{Q^2}{m^2} \bigg ) 
 \, \ln^{0, 1, 2} \bigg ( \frac{\mu^2}{m^2} \bigg ) 
 \,. 
\end{align}
The large logarithm $\ln(Q^2/m^2)$ is induced by the separation in rapidity of soft mass-shell fluctuations with the scaling  $(p^+,p^-,p_\perp) \sim(m,m,m)$ from collinear mass-shell fluctuations with $(p^+,p^-,p_\perp) \sim (m^2/Q,Q,m)$. It can not be eliminated by a choice of $\mu$ or summed by the RGE in $\mu$. This effect is directly related to virtual top quark loops which first appear at $\mathcal{O}(\alpha_s^2)$, and has been discussed in detail in Refs.~\cite{Gritschacher:2013pha,Pietrulewicz:2014qza} together with other subtleties concerning the incorporation of a massive quark in primary massless jet production in SCET. In Sec.~\ref{sec:direct} we will explicitly carry out the matching calculation for the ${\cal{O}}(\alpha_s^2 C_F T_F)$ correction with primary massive top quarks, and demonstrate how the amplitudes factorize into collinear and soft components which each involve a single rapidity scale. We show that this factorization is the same as that for massless external quarks, computed in Ref.~\cite{Pietrulewicz:2014qza}, up to a different constant term that appears in the collinear corrections. The direct computation of the SCET soft and collinear diagrams at $\mathcal{O}(\alpha_s^2 C_F T_F)$ can be performed elegantly by first computing the virtual correction for the radiation of a ``massive gluon" at one-loop and performing in a second step a dispersion integral.   In Sec.~\ref{sec:RRGandnum} we show how to resum the type of rapidity logarithm in \eq{rapidity_log} using the framework of the rapidity renormalization group established in Refs.~\cite{Chiu:2011qc,Chiu:2012ir}.
We also demonstrate that the residual scale dependence of $H_m$ on $\mu$ significantly decreases when employing the complete two-loop correction, and assess the impact of the rapidity logarithm. We conclude in Sec.~\ref{sec:conclusions}.

\section{Setup and Notation}
\label{sec:HmSetup}

As described in Refs.~\cite{Fleming:2007qr,Fleming:2007xt} for the description of the peak region we first match QCD onto SCET, and then SCET onto bHQET. The relevant current operators needed to define the hard functions in Eq.~(\ref{eq:dijetobservable}) are
\begin{align}
\label{eq:NewNotation}
{\cal{J}}_{\rm QCD}\hspace{8pt}& = \bar{\psi}(x) \Gamma_i^{\mu} \psi(x) \, ,
\nn \\
{\cal{J}}_{\rm SCET}& = \bar{\chi}_{n}  S_n^{\dagger}\Gamma_i^{\mu} S_{\bar{n}} \chi_{\bar{n}} \,,
\nn \\
{\cal{J}}_{\rm bHQET}&= \bar{h}_{v_+} W_n Y_n^{\dagger} \Gamma_i^{\mu}Y_{\bar{n}} W_{\bar{n}}^{\dagger} h_{v_-} \, , 
\end{align}
where $\Gamma_v^{\mu} = \gamma^{\mu}$ and $\Gamma_a^{\mu} = \gamma^{\mu} \gamma_5$. The jet fields $\chi_{n}=W_n^\dagger \xi_n$ and $\chi_{\bar{n}}=W_\bn^\dagger \xi_\bn$ describe the collinear radiation in SCET, and contain the massive collinear quarks $\xi_n$ and $\xi_\bn$~\cite{Leibovich:2003jd,Rothstein:2003wh} and Wilson lines $W_{n,\bn}$ where in position space $W^\dagger_n(x) ={\rm P}\exp\big(ig \int_0^\infty\! ds\, \bn\cdot A_n(\bn s+x)\big)$. The ultracollinear radiation in bHQET is described by the heavy quark fields $h_{v_{+,-}}$ and by $W_{n,\bar{n}}$. The wide-angle radiation in SCET is described by soft Wilson lines $S_{n,\bar{n}}$, where in position space $S^\dagger_n(x) ={\rm P}\exp\big(ig \int_0^\infty\! ds\, n\cdot A_s(n s+x)\big)$, and ultrasoft Wilson lines $Y_{n,\bar{n}}$ are the analogs with ultrasoft gluon fields in bHQET.  The difference between the SCET fields and bHQET fields is that SCET still contains soft and collinear fluctuations at the top mass scale, i.e.\ the SCET fields contain mass mode fluctuations which scale as $(p^+,p^-,p_\perp)\sim (m,m,m)$ and $(Q,m^2/Q,m)$ or $(m^2/Q,Q,m)$ which are absent in bHQET. This makes our EFT above the top mass scale an \SCETb type theory. There are six flavors in the ${\rm \overline{MS}}$ running coupling in QCD and SCET, and five flavors in bHQET.

The notation above differs from Ref.~\cite{Fleming:2007xt} which used a hybrid of \SCETa and \SCETb, where the current operator was written as 
\begin{align}
\label{eq:OldNotation}
 \widetilde{{\cal{J}}}_{\rm SCET} = \bar{\chi}_{n} Y_n^{\dagger} S_n^{\dagger}\Gamma_i^{\mu} S_{\bar{n}} Y_{\bar{n}} \chi_{\bar{n}} \,.
\end{align}
Here the Wilson lines $S_{n,\bar{n}}$ describe exclusively soft mass mode fluctuations and have ultrasoft zero-bin subtractions. In \eq{NewNotation} the SCET operator only describes soft fluctuations above and of order of the mass scale $m$, and not far below $m$. This simplifies the setup for the matching coefficient calculation, which in particular can be viewed as going from a six flavor theory to a five flavor theory.

The matching coefficients between these effective theories are defined by
\begin{align}
\label{eq:CQdefinition}
{{\cal J}}_{\rm QCD}^{(n_l+1)} &= C_Q^{(n_l+1)}\, {{\cal J}}_{\rm SCET}^{(n_l+1)}  
   \big[ 1 + {\cal O}(m/Q) \big ] \, ,\\
\label{eq:Cmdefinition}
{{\cal J}}_{\rm SCET}^{(n_l+1)} &=  C_m^{(n_f)}\,{{\cal J}}_{\rm bHQET}^{(n_l)} 
   \big [ 1 + {\cal O}(\hat s/m) \big ] \, .
\end{align}
Here both the currents and Wilson coefficients refer to the renormalized quantities. When we refer to the bare objects we will indicate this explicitly as e.g. in ${{\cal J}}_{\rm SCET}^{({\rm bare},n_l+1)} $. For all quantities we consider we use the renormalized coupling constant.  When we want to separate  the color structures of the matching coefficients we will do so in the following way:
\begin{align}
\label{eq:ColorStructure}
C_Q^{(n_l+1)}\hspace{2pt} &
  = 1 + C_Q^{(1,\,n_l+1)} +C_Q^{(C_F^2,\,n_l+1)} 
   + C_Q^{(C_F C_A, \, n_l+1 )} + C_m^{(C_F n_l T_F,\, n_l+1 )}
   + C_Q^{(C_F T_F , \, n_l+1 )} 
 , \nn \\
C_m^{(n_f)}\hspace{12pt} &= 1 + \underbrace{C_m^{(1,\,n_f)}}_{{\cal{O}}(\alpha_s)} \ +\   \underbrace{C_m^{(C_F^2,\,n_f)}  +  C_m^{(C_F C_A, \, n_f )} 
  + C_m^{(C_F n_l T_F,\, n_f )}  +  C_m^{(C_F T_F , \, n_f )} }_{{\cal{O}}(\alpha_s^2)} 
  .
\end{align}
In all the objects above the coupling is renormalized in the $\overline{\rm MS}$ scheme with the number of dynamical flavors, $n_f$, being either $n_l$ or $(n_l+1)$ as indicated by the superscript. Here $n_l$ is the number of light quarks, and the additional flavor indicates the heavy quark (here the top quark).  The choice for the number of flavors in each of the expressions above is motivated by the scales at which these objects live compared to the top mass.  Note that we have kept the number of flavors appearing in $C_m$ unspecified, as it can be expressed in either the $n_l$- or the $(n_l +1) $-flavor scheme. We will be explicit about which scheme we are using in the equations below.

The hard functions  in Eq.~(\ref{eq:dijetobservable}) are related to the Wilson coefficients via
\begin{align}
&H_Q(Q,\mu) = | C_Q|^2, \qquad H_m\bigg(m,\frac{Q}{m},\mu\bigg) = | C_m |^2 \, .
\end{align}
Here the dependence on $Q$ in the hard function $H_m$ appears due to the boost factor $Q/m$.

In Eq.~(\ref{eq:dijetobservable}) all the functions live at their respective scales and are evolved to a common scale $\mu_{\rm final}$ through renormalization group running. While the jet and the soft functions have convolution running~\cite{Fleming:2007xt}, the large logarithms of the hard matching coefficients are summed by multiplicative evolution factors,
\begin{align}  \label{eq:Hevol}
{H}_{\rm evol}(Q,m,\mu_{\rm final};\mu_Q,\mu_m,\nu_Q,\nu_m) \equiv & \, H^{(n_l+1)}_Q(Q,\mu_Q) \; U^{(n_l+1)}_{H_Q}(Q,\mu_Q, \mu_m) 
  \\
& \times H^{(n_l)}_m\bigg (m, \frac{Q}{m}, \mu_m ; \nu_Q,\nu_m \bigg) \;
 U^{(n_l)}_{v}\bigg (\frac{Q}{m},\mu_m, \mu_{\rm final}  \bigg) 
 \,, \nn
\end{align}
for $\mu_Q\simeq Q$, $\mu_m\simeq m$ and $\mu_{\rm final} < \mu_m$. On the LHS the dependence on $\mu_Q$ and $\mu_m$ only comes from higher order corrections when the objects in \eq{Hevol} are truncated at a given order in resummed perturbation theory. The same is true for the rapidity scales $\nu_Q$ and $\nu_m$, which are induced by the rapidity RGE that will be discussed further below and in \sec{RRG}. We will frequently drop these arguments that appear after the semicolon. 
The evolution factors here obey the RG equations 
\begin{align}
 \mu \frac{\df}{\df\mu} U^{(n_l+1)}_{H_Q}(Q,\mu_Q, \mu) 
   &= -\gamma^{(n_l+1)}_{H_Q}(Q,\mu) \, U^{(n_l+1)}_{H_Q}(Q,\mu_Q, \mu)  \,,
  \nn\\
  \mu \frac{\df}{\df\mu} 
    U^{(n_l)}_{v}\bigg (\frac{Q}{m},\mu, \mu_{\rm final}  \bigg)
   &= +\gamma^{(n_l)}_{v}\Big(\frac{Q}{m},\mu\Big)\, U^{(n_l)}_{v}\bigg (\frac{Q}{m},\mu, \mu_{\rm final}  \bigg)   \,,
\end{align}
where $\gamma^{(n_l)}_{v}$ is the anomalous dimension for the squared current in bHQET.

Eqs.~(\ref{eq:CQdefinition}) and~(\ref{eq:Cmdefinition}) suggest two different methods that one can use to calculate the ${\cal{O}}(\alpha_s^2)$ piece of $C_m$ or equivalently $H_m$: 
\\
\\
{\bf 1)} \textit{Indirect calculation using the known result for $C_Q$ and the matrix elements for the QCD and bHQET current operators in pure dimensional regularization:} 

Using Eq.~(\ref{eq:CQdefinition}) and~(\ref{eq:Cmdefinition}), and taking matrix elements of the operators with onshell top-quark states as in~\cite{Fleming:2007qr}, we have
\begin{align}
\langle {{\cal J}}_{\rm QCD}^{(n_l+1)} \rangle = C_Q^{(n_l+1)} \,  C_m^{(n_l)} \, \langle {{\cal J}}_{\rm bHQET}^{(n_l)} \rangle \, .
\end{align}
Using the relation between bare and renormalized bHQET currents
\begin{align}
\label{eq:bHQETamplitude}
\langle {{\cal J}}_{\rm bHQET}^{(n_l)} \rangle = Z_{\rm bHQET}^{(n_l)} \, \langle {{\cal J}}_{\rm bHQET}^{(\mathrm{ bare},\, n_l)}\rangle \, ,
\end{align}
we get
\begin{align}
\label{eq:Cmren}
 C_m^{(n_l)}  = \frac{ \langle {{\cal J}}_{\rm QCD}^{(n_l+1)} \rangle }
 { C_Q^{(n_l+1)}\, Z_{\rm bHQET}^{(n_l)}  \, \langle {{\cal J}}_{\rm bHQET}^{(\mathrm{ bare}, \, n_l)} \rangle } \, .
\end{align}
Note that the terms on the RHS involve objects with different flavor number schemes for the strong coupling, which must all be converted to $n_l$-flavors to get $C_m^{(n_l)}$.  Here we work in dimensional regularization for both UV and IR divergences and renormalize the quantities in the $\overline{\rm MS}$ scheme.  With this regulator we can use the known two loop result for the heavy form factor $\langle {{\cal J}}_{\rm QCD} \rangle$ given in Refs.~\cite{Bernreuther:2004ih, Gluza:2009yy}. The result for $C_Q$ is also known~\cite{Matsuura:1987wt,Matsuura:1988sm} in  $\overline{\rm MS}$, and the result for $Z_{\rm bHQET}^{(n_l)}$ can be determined by RG consistency as discussed below.  Loop graphs in bHQET factorize into ultrasoft and ultra-collinear contributions, and in general each involve at most a single dimensionful scale. The use of dimensional regularization for both the UV and IR, and employing onshell external quarks, imply that these loop corrections in bHQET are scaleless and vanish, such that $\langle {{\cal J}}_{\rm bHQET}^{(\mathrm{ bare},n_l)}\rangle =1$.  In general, the IR divergences in the QCD and bHQET matrix elements will match up, and the UV divergences in $\langle {{\cal J}}_{\rm bHQET}^{(\mathrm{ bare}, \, n_l)} \rangle$ are eliminated by the counterterm $Z_{\rm bHQET}^{(n_l)}$. In dimensional regularization with $1/\epsilon_{\rm IR} = 1/\epsilon_{\rm UV}$, this implies a cancellation of $1/\epsilon$ poles between $\langle {{\cal J}}_{\rm QCD}^{(n_l+1)} \rangle$ and $Z_{\rm bHQET}^{(n_l)}$. Thus we can use the simpler relation
\begin{align}
\label{eq:Cmbare}
  C_m^{(n_l)}=  \frac{\langle {{\cal J}}_{\rm QCD}^{(n_l+1)} \rangle}{  Z_{\rm bHQET}^{(n_l)} \,C_Q^{(n_l+1)} }\, .
\end{align}
\\
{\bf 2)} \textit{Direct calculation by matching the SCET and bHQET current operators:} 

Using Eq.~(\ref{eq:Cmdefinition}) we can also just directly compute the Wilson coefficient from a matching calculation, computing partonic matrix elements using the same IR regulator in SCET and bHQET,
\begin{align}
\label{eq:Cmmethodtwo}
C_m^{(n_l)}=\frac{ \langle {{\cal J}}_{\rm SCET}^{(n_l+1)}  \rangle }{ \langle {{\cal J}}_{\rm bHQET}^{(n_l)}\rangle }
 \equiv \frac{F_{\rm SCET}^{(n_l+1)} }{F_{\rm bHQET}^{(n_l)} } \, .
\end{align} 
These matrix elements are form factors in the respective theories which we denote by $F$. 
We will use the same notation for the color structures in the perturbative expansion of $F_{\rm SCET}$ and $F_{\rm bHQET}$ as in \eq{ColorStructure}. We define the relation between bare and renormalized SCET currents by
\begin{align}
  \label{eq:SCETamplitude}
\langle {{\cal J}}_{\rm SCET}^{(n_l+1)} \rangle = Z_{\rm SCET}^{(n_l+1)} \, \langle {{\cal J}}_{\rm SCET}^{(\mathrm{ bare},\, n_l+1)}\rangle \,.
\end{align}
As usual the bare currents are $\mu$-independent, so from Eqs.~(\ref{eq:bHQETamplitude}),~(\ref{eq:Cmmethodtwo}) and~(\ref{eq:SCETamplitude}) the $\mu$-RG equation for $C_m^{(n_l)}$ can be written as
\begin{align} \label{eq:muRGE}
  \mu \,\frac{\df}{\df \mu} \,\ln \,C_m^{(n_l)} 
  = \Big[\gamma^{(n_l+1)}_{\rm SCET}(Q,\mu) -\gamma^{(n_l)}_{\rm bHQET}\Big(\frac{Q}{m},\mu\Big) \Big]
  (\alpha_s^{(n_l)}) 
  \equiv \gamma^{C_m}_\mu(Q,m,\mu) \,,
\end{align}
where the current anomalous dimensions are computed order-by-order from the counterterms in the standard fashion 
\begin{align}  \label{eq:gammamufromZ}
  \gamma^{(n_l+1)}_{\rm SCET}(Q,\mu)
     &= \mu \,\frac{\df}{\df \mu} \ln \,Z_{\rm SCET}^{(n_l+1)} \,,
  & \gamma^{(n_l)}_{\rm bHQET}\Big(\frac{Q}{m},\mu\Big)
     &= \mu \,\frac{\df}{\df \mu} \ln \,Z_{\rm bHQET}^{(n_l)} \,.
\end{align}
The anomalous dimension for the SCET current is known to 3-loop order~\cite{Moch:2005id}. Up to two loops the result reads
\begin{align}  \label{eq:gammaSCET}
\gamma^{(n_l+1)}_{\rm SCET}(Q,\mu) &= \frac{\alpha_s^{(n_l+1)}(\mu)C_F}{4\pi}\big[\!-4 L_Q +6\big] +\bigg ( \frac{\alpha_s^{(n_l+1)}(\mu)}{4 \pi} \bigg)^{\!2} \bigg\{C_F^2 \big[3 - 4 \pi^2 + 48 \zeta_3\big] \nn \\
&  \quad +C_F C_A\bigg[-\bigg ( \frac{268}{9} -\frac{4\pi^2}{3}   \bigg )L_Q+ \frac{961}{27} + \frac{11\pi^2}{3} - 52 \zeta_3\bigg]  \nn \\
    &
  \quad+ (n_l+1) C_F T_F\bigg[\frac{80}{9}L_Q-\frac{260}{27}-\frac{4\pi^2}{3}\bigg]\bigg\} \,,
\end{align}
where $L_Q=\ln[(-Q^2-i0)/\mu^2]$. 
The bHQET anomalous dimension can be derived using one of the consistency relations~\cite{Fleming:2007xt} for the factorization theorem in \eq{dijetobservable}:
\begin{align}\label{eq:consistency_bHQET}
 \gamma_v = \gamma_{\rm bHQET} +\gamma^*_{\rm bHQET}
          = 2\gamma_{\Jht} +2\gamma_{S} \,,
\end{align}
where $\gamma_S$ indicates the soft function anomalous dimension for one hemisphere. Using the results for $\gamma_{\Jht}$ given in Eq.~(41) of Ref.~\cite{Jain:2008gb}  and for $\gamma_{S}$ given in Eq.~(19) of Ref.~\cite{Hoang:2008fs} (which can be derived via consistency from the two-loop jet function anomalous dimension~\cite{Becher:2006qw}) we find
\begin{align}  \label{eq:gammabHQET}
  \gamma_{\rm bHQET}\Big(\frac{Q}{m},\mu\Big)
    &
  = \frac{\alpha_s^{(n_l)}(\mu)C_F}{4\pi}\big[\!-4 L +4\big]   +\bigg ( \frac{\alpha_s^{(n_l)}(\mu)}{4 \pi} \bigg)^{\!2} \bigg\{
 n_l C_F T_F\bigg[\frac{80}{9}L-\frac{80}{9}\bigg]
 \nn \\
    &\ \ 
  \quad+ C_F C_A\bigg[-\bigg (\frac{268}{9}- \frac{4\pi^2}{3}    \bigg )L+ \frac{196}{9} - \frac{4\pi^2}{3} + 8 \zeta_3\bigg] \bigg\} +\mathcal{O}(\alpha_s^3)\,,
\end{align}
where $L=\ln[(-Q^2-i0)/m^2]$. Expanding the recently calculated anomalous dimension in HQET at $\mathcal{O}(\alpha_s^3)$ \cite{Grozin:2014hna,Grozin:2015kna} we extract in appendix~B also the three-loop coefficient, which has -- to our knowledge -- not yet been displayed in literature.

As mentioned above, the two-loop expression of $C_m$ contains large logarithms of the form $\alpha_s^2 C_F T_F\, \ln(-m^2/Q^2) \sim {\cal{O}}(\alpha_s)$ which cannot be resummed using the RGE in $\mu$. They are rapidity logarithms and originate from a separation of the soft and collinear mass modes which have the same invariant mass but different rapidity. These rapidity logarithms only appear inside $H_m$, and not for the other soft, jet, and hard functions in \eq{dijetobservable}. Our focus here will be on the leading rapidity logarithms, which start contributing with the $\mathcal{O}(\alpha_s^2 C_F T_F)$ piece.  The latter comes from virtual top quark loops, and hence we only need to compute the correction $ F_{\rm SCET}^{(C_F T_F , \,n_l+1)}$, while the bHQET graphs give no contribution for this color structure, i.e. $F_{\rm bHQET}^{(C_FT_F,n_l)}=0$.

To set up the stage for rapidity resummation we can factorize the current operators and its matrix elements into products of soft and collinear diagrams, 
\begin{align} \label{eq:Jsplit}
   \langle {{\cal J}}_{\mathrm{ SCET}}^{(n_l+1)}  \rangle 
   &= \langle {{\cal J}}_{\mathrm{ SCET}}^{(n_l+1)}  \rangle_{n} \
     \langle {{\cal J}}_{\mathrm{ SCET}}^{(n_l+1)}  \rangle_{s} \
     \langle {{\cal J}}_{\mathrm{ SCET}}^{(n_l+1)}  \rangle_{\bn}   
  \,, \nn\\
  \langle {{\cal J}}_{\mathrm{bHQET}}^{(n_l+1)}  \rangle 
   &= \langle {{\cal J}}_{\mathrm{bHQET}}^{(n_l+1)}  \rangle_{n} \
     \langle {{\cal J}}_{\mathrm{bHQET}}^{(n_l+1)}  \rangle_{s} \
     \langle {{\cal J}}_{\mathrm{bHQET}}^{(n_l+1)}  \rangle_{\bn} 
  \,,
\end{align}
where the $\{n,s,\bn\}$ labels in bHQET indicate $n$-ucollinear, ultrasoft, and $\bn$-ucollinear contributions respectively. Note that in order to split up these corrections we must choose an IR regulator which preserves the \SCETb nature of the theory. We will regulate the IR divergences using a gluon mass $\Lambda$, which thus differs from the use of pure dimensional regularization discussed above for method 1. In \SCETb the individual soft and collinear diagrams have rapidity divergences, and using the regulator of Refs.~\cite{Chiu:2011qc,Chiu:2012ir} the coefficients will depend on a rapidity renormalization scale $\nu$. Thus Eq.~(\ref{eq:Cmmethodtwo}) can be decomposed into individual contributions involving $n$-collinear, $\bar{n}$-collinear, and soft amplitudes,
\begin{align} \label{eq:Cmsplit}
C_{m, \, i}^{(n_l)} 
  =\frac{ \langle {{\cal J}}_{\mathrm{ SCET}}^{(n_l+1)}  \rangle_{i} }{ \langle {{\cal J}}_{\mathrm{bHQET}}^{(n_l)} \rangle_i } \,, \qquad i = n,\bar{n},s \, .
\end{align}
This leads to
\begin{align} \label{eq:Hmfactorized} 
  C_m^{(n_l)}\bigg(m,\frac{Q}{m},\mu\bigg)
   &= C_{m,n}^{(n_l)}\bigg(m,\mu,\frac{\nu}{Q}\bigg)\,
      C_{m,s}^{(n_l)}\bigg(m,\mu,\frac{\nu}{m}\bigg)\,
      C_{m,\bn}^{(n_l)}\bigg(m,\mu,\frac{\nu}{Q}\bigg)
  \,, 
\end{align}
where we included the dependence on scales and renormalization parameters.
Thus we see that the logarithmic dependence on the $Q/m$ boost variable is factorized by the rapidity regularization parameter $\nu$ into collinear factors that depend on $Q$ and a soft factor which does not. 
To sum the rapidity logarithms we can follow the standard approach of matching and running.  

We define hard functions $H_{m, i}^{(n_l)}=\Big|C_{m,i}^{(n_l)}\Big|^2$.  
The individual Wilson coefficient and hard functions obey related RG equations, 
\begin{align}
\label{eq:GammaNus}
  \nu \frac{\df}{\df\nu} C_{m,i}^{(n_l)} 
   = \gamma_{\nu,i}^{C_m}\,  C_{m,i}^{(n_l)} 
  \,,  \qquad
   \nu \frac{\df}{\df\nu} H_{m,i}^{(n_l)} 
   = \gamma_{\nu,i}^{H_m}\,  H_{m,i}^{(n_l)} 
  \,,  \qquad
\gamma^{H_m}_{\nu,\, i} 
   = \gamma^{C_m}_{\nu, \, i}
    +  \big(\gamma^{C_m}_{\nu, \, i}\big)^* \,.
\end{align}
The $\nu$-anomalous dimensions appearing here can be computed directly from the SCET and bHQET counterterms and depend only on $m$ and $\mu$.  Taking \eqs{bHQETamplitude}{SCETamplitude} and introducing individual counterterm factors for each of the collinear and soft component amplitudes, noting that the bare coefficients are $\nu$-independent, and using Eq.~(\ref{eq:Cmsplit}) we get 
\begin{align}  \label{eq:gammaCm}
 \gamma^{C_m}_{\nu, \, i}(m,\mu)= \nu \frac{\df}{\df \nu} \ln \, C_{m, \, i}^{(n_l)} 
 &=  \nu \frac{\df}{\df \nu} \ln \,  \langle {{\cal J}}_{\mathrm{SCET}}^{(n_l+1)}  \rangle_i  
 \,-  \nu \frac{\df}{\df \nu} \ln \, \langle {{\cal J}}_{\mathrm{bHQET}}^{(n_l)} \rangle_i 
 \nn\\
&=  \nu \frac{\df}{\df \nu} \ln\, Z_{{\rm SCET},i}^{(n_l+1)} - 
   \nu \frac{\df}{\df \nu} \ln\, Z_{{\rm bHQET},i}^{(n_l)} 
  \, , \qquad i=n,\bar{n},s \, .  
\end{align}
As we will see in detail below, individual contributions on the right hand side of Eq.~(\ref{eq:gammaCm}) contain IR divergences, but they will always cancel to leave an IR finite result for the $\gamma^{C_m}_{\nu, \, i}$, when we fully expand in either the $n_l$-flavor or $(n_l+1)$-flavor scheme for the strong coupling.

\section{Two Loop Determination of $H_m$ from QCD heavy form factor}
\label{sec:Hm}

In this section we use the first method outlined in \sec{HmSetup} to determine the bHQET matching coefficient, $C_m$ at two loops. From Eq.~(\ref{eq:Cmbare}) the ingredients we need are the UV renormalized QCD two-loop heavy quark form factor, $\langle {{\cal J}}_{\rm QCD}^{(n_l+1)} \rangle$, in dimensional regularization and the SCET matching coefficient, $C_Q^{(n_l+1)}$. In the following we abbreviate the appearing logarithms as
\begin{align} \label{eq:logs}
 L &= \ln \bigg ( \frac{-Q^2- i 0}{m^2} \bigg), 
 & L_{m} & = \ln \bigg ( \frac{m^2}{\mu^2} \bigg), 
 & L_Q & = \ln \, \bigg ( \frac{- Q^2 -i 0 }{\mu^2 } \bigg ) \,.
\end{align}

From Refs.~\cite{Bernreuther:2004ih, Gluza:2009yy} we extract the renormalized two loop QCD heavy quark form factor result in the high energy limit, $Q^2 \gg m^2$, evaluated at an arbitrary scale $\mu \gtrsim m$, abbreviating $\alpha_s^{(n_l+1)} \equiv\alpha_s^{(n_l+1)}(\mu)$,~\footnote{Note that in Ref.~\cite{Bernreuther:2004ih} the counterterm for the renormalization of the coupling constant contains an extra factor $\Gamma(1+\epsilon)$, so that also additional finite terms are subtracted compared to the conventional $\MS$ renormalization.} 
\begin{align}
 F^{(n_l+1)}_{\rm QCD} &=  1+  \frac{\alpha_s^{(n_l+1)} C_F}{4 \pi}  \, 
  \bigg\{ \frac{2 L -2}{\epsilon} -L^2 - (2\lmu-3) L+2\lmu-4 + \frac{\pi^2}{3} + \epsilon \bigg [ \frac{L^3}{3} +\bigg(\!\lmu-\frac{3}{2}\bigg) L^2 \nn \\
& \quad  + \bigg(\lmu^2-3\lmu+8-\frac{\pi^2}{6}\bigg)L-\lmu^2+ \bigg (4-\frac{\pi^2}{3}\bigg)\lmu - 8 + \frac{\pi^2}{3}+ 4 \zeta_3 \bigg ]  + {\cal O}(\epsilon^2)  \bigg\} \nn \\
&  +\bigg ( \frac{\alpha_s^{(n_l+1)}}{4 \pi} \bigg)^2 C_F^2 \,  \bigg \{ \frac{1}{\epsilon^2} \big[2 L^2 \!-\! 4 L + 2\big] + \frac{1}{\epsilon} \bigg[\!-2L^3 \!-\! (4\lmu\!-\! 8)L^2 + \bigg(8 \lmu \!- 14 
 + \frac{2\pi^2}{3}\bigg) L   \nn \\
 & \quad - 4\lmu+ 8 - \frac{2\pi^2}{3}\bigg]+ \frac{7}{6} L^4 +\bigg(4 \lmu-\frac{20}{3}  \bigg) L^3 + \bigg(4\lmu^2-16\lmu+\frac{55}{2} - \frac{2\pi^2}{3} \bigg) L^2 \nn \\
&\quad  - \bigg (8\lmu^2 -\bigg(28-\frac{4\pi^2}{3}\bigg) \lmu +\frac{85}{2}-32 \zeta_3 \bigg ) L+4\lmu^2 - \bigg(16 - \frac{4\pi^2}{3}\bigg) \lmu + 46 + \frac{13\pi^2}{2} \nn \\
&\quad -44 \zeta_3 -8 \pi^2 \,\ln \,2 - \frac{59\pi^4}{90} + {\cal O}(\epsilon) \bigg \} \nn \\
& +\bigg ( \frac{\alpha_s^{(n_l+1)}}{4 \pi} \bigg)^2 C_F C_A  \, \bigg \{ \frac{1}{\epsilon^2} \bigg [\! -\frac{11}{3} L \!+\! \frac{11}{3} \bigg] + \frac{1}{\epsilon} \bigg [ \bigg( \frac{67}{9} - \frac{\pi^2}{3} \bigg ) L -\frac{49}{9} +  \frac{\pi^2}{3} - 2\zeta_3 \bigg]+ \frac{11}{9} L^3 \nn \\
&\quad  + \bigg ( \frac{11 }{3}\lmu-\frac{233}{18}+ \frac{\pi^2}{3} \bigg) L^2  + \bigg (\frac{11}{3} \lmu^2 - \bigg (\frac{233}{9} - \frac{2\pi^2}{3} \bigg) \lmu +\frac{2545}{54} + \frac{11\pi^2}{9} - 26 \zeta_3 \bigg) L\nn \\
& \quad -\frac{11}{3} \lmu^2+ \bigg (\frac{230}{9} - \frac{17\pi^2}{9} + 4 \zeta_3 \bigg) \lmu- \frac{1595}{27} - \frac{7\pi^2}{54} + \frac{134}{3} \zeta_3 + 4 \pi^2 \, \ln \, 2 - \frac{\pi^4}{60}+ {\cal O}(\epsilon) \bigg \} \nn \\
&  +\bigg ( \frac{\alpha_s^{(n_l+1)}}{4 \pi} \bigg)^2 C_F n_l T_F \, \bigg \{ \frac{1}{\epsilon^2} \bigg[ \frac{4}{3} L - \frac{4}{3} \bigg] + \frac{1}{\epsilon} \bigg [\! -\frac{20}{9} L + \frac{20}{9} \bigg] - \frac{4}{9} L^3 - \bigg(\frac{4}{3} \lmu -\frac{38}{9}  \bigg) L^2  \nn \\
&\quad- \bigg (\frac{4}{3} \lmu^2 - \frac{76}{9} \lmu+\frac{418}{27} + \frac{4\pi^2}{9} \bigg ) L+ \frac{4}{3} \lmu^2  - \bigg (\frac{88}{9} -\frac{4\pi^2}{9} \bigg) \lmu+ \frac{424}{27} - \frac{14\pi^2}{27} \nn \\
& \quad - \frac{16}{3} \zeta_3 + {\cal O}(\epsilon) \bigg \} \nn \\
&  +\bigg ( \frac{\alpha_s^{(n_l+1)}}{4 \pi} \bigg)^2 C_F T_F \,\bigg \{ \frac{1}{\epsilon} \bigg[ \frac{8}{3} \lmu \, L-\frac{8}{3} \lmu \bigg]  -\frac{4}{9} L^3 - \bigg(\frac{4}{3} \lmu-\frac{38}{9} \bigg)L^2  - \bigg (\! 4 \lmu^2 -4\lmu\nn \\
&\quad +\frac{530}{27}+\frac{2\pi^2}{3} \bigg) L +4 \lmu^2 - \bigg( \frac{16}{3} - \frac{4\pi^2}{9}\bigg) \lmu+ \frac{1532}{27} - \frac{4 \pi^2}{9} + {\cal O} (\epsilon)\bigg \}   \, .
\end{align}
Note that we keep the ${\cal O } (\epsilon) $ part of the one loop piece in $F_{\rm QCD}^{(1,n_l+1)}$ since it yields a contribution when considering the cross terms in the expansion of the ratio in \eq{Cmbare}. (One can avoid considering these cross terms and obtain the same answer by taking the logarithm of \eq{Cmbare}.) We remark that in these expressions the pole mass scheme has been used for the top quark mass $m$.

The other ingredient we need is the well known two-loop expression for $C_Q$, widely used in the SCET literature, and obtained with the aid of  the massless form factor calculation of Refs.~\cite{Matsuura:1987wt,Matsuura:1988sm}, 
\begin{align}
\label{eq:CQren}
C_Q^{(n_l+1)} &= 1 + \frac{\alpha_s^{(n_l+1)}(\mu) C_F}{4 \pi} \, \bigg \{\! - \lqp^2  +3 \lqp -8 +\frac{\pi^2}{6} \bigg \} 
 \nn \\
& +\bigg ( \frac{\alpha_s^{(n_l+1)}(\mu)}{4 \pi} \bigg)^2 C_F^2 \, \bigg \{ \frac{1}{2} \lqp^4-3\lqp^3+\bigg (\frac{25}{2}-\frac{\pi^2}{6} \bigg) \lqp^2 -\bigg (\frac{45}{2} +\frac{3 \pi^2}{2} - 24 \zeta_3 \bigg ) \lqp
  \nn \\
&\hspace{100pt}
 + \frac{255}{8}+ \frac{7 \pi^2}{2}   - 30 \zeta_3- \frac{83 \pi^4}{360}  \bigg \} 
 \nn \\
&+ \bigg  ( \frac{\alpha_s^{(n_l+1)}(\mu)}{4 \pi} \bigg)^2 C_A C_F  \, \bigg \{ \frac{11}{9}\lqp^3- \bigg( \frac{233}{18}- \frac{\pi^2}{3} \bigg) \lqp^2 + \bigg ( \frac{2545}{54}+\frac{11 \pi^2}{9} -26 \zeta_3 \bigg ) \lqp 
 \nn \\
& \hspace{100pt}
 -\frac{51157}{648} - \frac{337 \pi^2}{108} + \frac{313 \zeta_3}{9} + \frac{11 \pi^4}{45}\bigg \} 
  \nn \\
&+ \bigg ( \frac{\alpha_s^{(n_l+1)}(\mu)}{4 \pi} \bigg)^2 C_F  T_F (n_l+1) \, \bigg \{ \!-\frac{4}{9}\lqp^3 + \frac{38}{9}\lqp^2 - \bigg ( \frac{418}{27}+\frac{4 \pi^2}{9} \bigg ) \lqp
 \nn\\
 & \hspace{100pt}
+\frac{4085}{162}  + \frac{23 \pi^2}{27} + \frac{4 \zeta_3}{9} \bigg \}
   \, .
\end{align}

The remaining quantities in Eq.~(\ref{eq:Cmbare}) are the coefficient $C_m^{(n_l)}$ we wish to determine, and the counterterm $Z_{\rm bHQET}^{(n_l)}$. The contributions to these two quantities can be easily distinguished since $Z_{\rm bHQET}^{(n_l)}$ only has terms with powers of $1/\epsilon$, whereas $C_m^{(n_l)}$ is given by the finite $\mathcal{O}(\epsilon^0)$ contribution. Therefore, it is straightforward to distinguish these two quantities unambiguously. Since we wish to determine these with $n_l$ active flavors, we must convert the strong coupling in  $\langle {{\cal J}}_{\rm QCD}^{( n_l + 1)} \rangle $ and $C_Q^{(n_l+1)}$ to the $n_l$-flavor scheme using the decoupling relation
\begin{align}
\label{eq:decouplingepsilon}
\alpha_s^{(n_l+1)}(\mu)
 &= \alpha_s^{(n_l)}(\mu) \Bigg\{ 1 + \alpha_s^{(n_l)}(\mu)  \bigg[\Pi(m^2,0) - \frac{\, \alpha_s^{(n_l)} (\mu) T_F}{3\pi} \frac{1}{\epsilon}\bigg] +\mathcal{O}(\alpha_s^2)\Bigg\} \, ,
\end{align}
where the one-loop vacuum polarization at zero momentum transfer for a massive quark pair is given by
\begin{align} \label{eq:pi0}
  \Pi(m^2,0) &=  \frac{\alpha_s(\mu) T_F}{3\pi} \bigg(\frac{\mu^2 e^{\gamma_E}}{m^2}\bigg)^{\epsilon} 
  \Gamma(\epsilon) = \frac{\alpha_s(\mu) T_F}{3\pi}\bigg[\frac{1}{\epsilon}-L_m+\epsilon\bigg(\frac{1}{2}L_m^2+\frac{\pi^2}{12}\bigg)+\mathcal{O}(\epsilon^2)\bigg] \, .
\end{align}
We need to keep terms up to ${\cal{O}}(\epsilon)$ in \eq{decouplingepsilon} since they contribute in the dimensional regularization scheme we are using when multiplying ${\cal{O}}(\alpha_s/\epsilon)$ IR divergent terms in Eq.~(\ref{eq:Cmbare}).  Using these results in Eq.~(\ref{eq:Cmbare}) we find the following expression for $Z_{\rm bHQET}^{(n_l)}$,
\begin{align}
\label{eq:ZbHQET5f}
 Z_{\rm bHQET}^{(n_l)} &= 
 1 +  \frac{\alpha_s^{(n_l)}(\mu)C_F}{4 \pi}  \, \frac{1}{\epsilon}\, \big( 2 L -2 \big)
 +\bigg ( \frac{\alpha_s^{(n_l)}(\mu)}{4 \pi} \bigg)^2 C_F^2 \,\frac{1}{\epsilon^2} \,\big( 2 L^2 -4L + 2\big) \nn \\
& \quad+\bigg ( \frac{\alpha_s^{(n_l)}(\mu)}{4 \pi} \bigg)^2 C_F C_A \bigg \{\frac{1}{\epsilon^2} \bigg[- \frac{11}{3}L+ \frac{11}{3} \bigg] + \frac{1}{\epsilon} \bigg[\bigg ( \frac{67}{9} - \frac{\pi^2}{3} \bigg )L - \frac{49}{9} + \frac{\pi^2}{3} - 2 \zeta_3\bigg]\bigg\} \nn \\
& \quad +\bigg ( \frac{\alpha_s^{(n_l)}(\mu)}{4 \pi} \bigg)^2 C_F n_l T_F \bigg \{   \frac{1}{\epsilon^2} \bigg[\frac{4}{3} L -\frac{4}{3}\bigg]  + \frac{1}{\epsilon} \bigg[- \frac{20}{9} L+ \frac{20}{9}\bigg]\bigg\} 
 \, . 
\end{align}
This result can also be extracted from earlier literature using the consistency relation for RG running between $H_m$, and the soft and the jet functions in Eq.~(\ref{eq:dijetobservable}). In particular, the $1/\epsilon^2$ terms in  \eq{ZbHQET5f} are given by a term involving the lowest order $\beta$-function, and the square of the one-loop result (due to non-abelian exponentiation), while the $1/\epsilon$ terms are directly related to the two-loop anomalous dimension given in \eq{gammabHQET}. This provides a key cross-check for $Z_{\rm bHQET}^{(n_l)}$ and hence for our result below for $C_m^{(n_l)}$.

After cancellation of the $1/\epsilon$ and $1/\epsilon^2$ terms in \eq{Cmbare} with the help of $Z_{\rm bHQET}^{(n_l)}$, the remaining ${\cal O}(\epsilon^0)$ terms give the desired result for $C_m^{(n_l)}$.  With the top-mass in the pole scheme we find
\begin{align}
\label{eq:Cm5fren}
 & C_m^{(n_l)}\Big(m,\frac{Q}{m},\mu\Big)
 = 1+\frac{\alpha_s^{(n_l)}(\mu)C_F }{4 \pi} \, \bigg (  \lmu^2 -\lmu + 4 +\frac{\pi^2}{6} \bigg ) \nn \\
&\qquad
+\bigg ( \frac{\alpha_s^{(n_l)}(\mu)}{4 \pi} \bigg)^2 C_F^2 \, \bigg \{  \frac{1}{2}\lmu^4-\lmu^3+ \bigg( \frac{9}{2}+\frac{\pi^2}{6} \bigg) \lmu^2- \bigg ( \frac{11}{2} -\frac{11 \pi^2}{6} + 24 \zeta_3 \bigg) \lmu  \nn \\
&\qquad\qquad
 +\frac{241}{8}+ \frac{13 \pi^2}{3} - 8 \pi^2 \log 2 - 6 \zeta_3-\frac{163 \pi^4}{360}   \bigg \}  \nn \\
&\qquad
+\bigg ( \frac{\alpha_s^{(n_l)}(\mu)}{4 \pi} \bigg)^2 C_A C_F \, \bigg \{ -\frac{11}{9} \lmu^3 + \bigg ( \frac{167}{18}- \frac{\pi^2}{3} \bigg) \lmu^2  - \bigg ( \frac{1165}{54}+ \frac{28 \pi^2}{9} - 30 \zeta_3 \bigg) \lmu  \nn \\
&\qquad\qquad  
 + \frac{12877}{648} + \frac{323 \pi^2}{108} + 4 \pi^2 \log 2 + \frac{89 \zeta_3}{9} - \frac{47 \pi^4 }{180}    \bigg \} \nn \\
&\qquad
 +\bigg ( \frac{\alpha_s^{(n_l)}(\mu)}{4 \pi} \bigg)^2 C_F n_l T_F \, \bigg \{  \frac{4}{9}\lmu^3 - \frac{26}{9}\lmu^2  + \bigg ( \frac{154}{27} + \frac{8 \pi^2}{9} \bigg) \lmu   -\frac{1541}{162} - \frac{37 \pi^2}{27} -\frac{52\zeta_3}{9}  \bigg \} \nn \\
&\qquad
 +\bigg ( \frac{\alpha_s^{(n_l)}(\mu)}{4 \pi} \bigg)^2 C_F T_F \, \bigg \{  -\frac{8}{9} \lmu^3 - \frac{2}{9} \lmu^2+\bigg ( \frac{130}{27}+ \frac{2 \pi^2}{3}  \bigg) \lmu
 +\frac{5107}{162} -\frac{41 \pi^2 }{27} - \frac{4 \zeta_3}{9}
 \nn \\
&\qquad\qquad 
  - \bigg (  \frac{4}{3}  \lmu^2+ \frac{40}{9}  \lmu+\frac{112}{27} \bigg) \ln \, \bigg ( \frac{-Q^2\minus i0}{m^2}\bigg)    
    \bigg \}  \, ,
\end{align}
Finally we arrive at the main result of this section - the result for $H_m = |C_m|^2$ in the $n_l$-flavor scheme with the top-mass in the pole scheme ($\alpha_s^{(n_l)} \equiv \alpha_s^{(n_l)}(\mu)$)
\begin{align}
\label{eq:Hm5f}
H_m^{(n_l)}\Big(m,\frac{Q}{m},\mu\Big) 
&=1+ \frac{\alpha_s^{(n_l)}(\mu)}{4 \pi} C_F \, \bigg ( 2 \lmu^2- 2 \lmu + 8 +\frac{\pi^2}{3} \bigg ) \nn \\
&\hspace{-2cm}
+\bigg ( \frac{\alpha_s^{(n_l)}(\mu)}{4 \pi} \bigg)^2 C_F^2 \, \bigg \{ 2 \lmu^4 - 4 \lmu^3 + \bigg ( 18 + \frac{2 \pi^2}{3} \bigg ) \lmu^2- \bigg ( 19 -\frac{10 \pi^2}{3} + 48 \zeta_3 \bigg ) \lmu \nn \\
&\hspace{-2cm}
\qquad\qquad  + \frac{305}{4} + 10 \pi^2  - 16 \pi^2 \log 2 - 12 \zeta_3-\frac{79 \pi^4}{90}  \bigg \}  \nn \\
&\hspace{-2cm}
+\bigg(\! \frac{\alpha_s^{(n_l)}(\mu)}{4 \pi}\! \bigg)^2 C_A C_F \, \bigg \{ -\frac{22}{9} \lmu^3+ \bigg ( \frac{167}{9} - \frac{2\pi^2}{3} \bigg) \lmu^2  - \bigg ( \frac{1165}{27}+ \frac{56 \pi^2}{9} - 60 \zeta_3 \bigg) \lmu   \nn \\
&\hspace{-2cm}
\qquad\qquad    + \frac{12877}{324} + \frac{323 \pi^2}{54} + 8 \pi^2 \log 2 + \frac{178 \zeta_3}{9}- \frac{47 \pi^4 }{90} \bigg \} \nn \\
&\hspace{-2cm}
+\bigg(\! \frac{\alpha_s^{(n_l)}(\mu)}{4 \pi} \!\bigg)^2 C_F n_l T_F \, \bigg \{  \frac{8}{9}\lmu^3 - \frac{52}{9}\lmu^2 + \bigg ( \frac{308}{27} + \frac{16 \pi^2}{9} \bigg) \lmu  -\frac{1541}{81} - \frac{74 \pi^2}{27}   -\frac{104\zeta_3}{9}  \bigg \}\nn \\
&\hspace{-2cm}
+\bigg(\!\frac{\alpha_s^{(n_l)}(\mu)}{4 \pi} \!\bigg)^2 C_F T_F \, \bigg \{-\frac{16}{9} \lmu^3 - \frac{4}{9} \lmu^2   +\bigg ( \frac{260}{27}+ \frac{4 \pi^2}{3}  \bigg) \lmu  
+ \frac{5107}{81} -\frac{82 \pi^2 }{27}- \frac{8 \zeta_3}{9}     
  \nn \\
&\hspace{-2cm}
\qquad\qquad 
- \bigg(  \frac{8}{3}  \lmu^2 + \frac{80}{9}  \lmu  +\frac{224}{27}   \bigg ) \, \ln \bigg ( \frac{Q^2}{m^2} \bigg) 
\bigg \} \, .
\end{align}
As anticipated, all of the logarithms in this expression are minimized for $\mu\simeq m$, except for the contributions in the last line that involve the rapidity logarithm $\alpha_s^2 C_F T_F\, \ln(Q^2/m^2)$. To understand the origin of this type of logarithm in the context of the renormalization group requires a further factorization of $H_m^{(n_l)}$ into soft and collinear pieces, as in \eq{Hmfactorized}.  In the next section we will carry out an independent calculation of the ${\cal O}(\alpha_s^2C_F T_F)$ terms in $H_m^{(n_l)}$. This sets up the rapidity renormalization group analysis of this term, which can be found in \sec{RRG}.  In \sec{nums} we present the result for $H_{\bar m}^{(n_l+1)}$  with the top mass renormalized in the $\MS$ scheme.

\section{Direct Computation of the ${\cal O}(\alpha_s^2 C_F T_F)$ Result} 
\label{sec:direct}
\subsection{Ingredients for the Calculation}
\label{sec:DirectSketch}

In this section we perform a direct computation of the $\alpha_s^2 C_F T_F$ piece of the matching coefficient $C_m(m,Q/m,\mu)$ due to massive quark loops using the second method from \sec{HmSetup}. We carry out the calculation in analogy to Refs.~\cite{Gritschacher:2013pha,Pietrulewicz:2014qza}, where the corresponding contribution to the matching coefficient at the mass scale for massless external quarks (in the following called ``primary") was computed. In this section we extend the calculation to the case of primary massive quarks.

Starting from Eq.~(\ref{eq:Cmmethodtwo}) we note that for the $\alpha_s^2 C_F T_F$ massive quark term, the bHQET graphs expressed in the usual $n_l$-flavor scheme do not give any contribution. The SCET graphs do contribute, and should be expressed in the same scheme for the strong coupling. Using the decoupling relation in Eq.~(\ref{eq:decouplingepsilon})  we obtain in the notation of Eq.~(\ref{eq:ColorStructure})
\begin{align}
\label{eq:CmCFTF5f}
C_m^{(C_F T_F, \, n_l)}\Big(m,\frac{Q}{m},\mu\Big) 
 &= \bigg[ F_{\rm SCET}^{(C_F T_F , \,n_l+1)}(Q,m,\Lambda,\mu) 
 \\
 &\qquad - \frac{ \alpha_s^{(n_l)} (\mu) T_F}{3\pi}\, \ln \bigg (\frac{m^2}{\mu^2} \bigg) \,  F_{\rm SCET}^{(1,n_l+1)}(Q,m,\Lambda,\mu)
 \bigg]_{ \alpha_s^{(n_l+1)} \to\, \alpha_s^{(n_l)} } 
 \,. \nn 
\end{align}
The second term on the right hand side accounts for the coupling conversion of the SCET form factor from $(n_l+1)$ to $n_l$ flavors.\footnote{Note that the subscript ``$\alpha_s^{(n_l+1)}\to \alpha_s^{(n_l)}$'' used here and elsewhere stands for the plain replacement of the couplings and does not involve any expansion based on \eq{decouplingepsilon}.} As discussed in detail below, we will use a massive gluon as an IR regulator $\Lambda$, such that ${\cal{O}}(\epsilon)$ terms in the coupling conversion in \eq{decouplingepsilon} can be dropped. For the remainder of this section we will drop the superscript $(n_l+1)$ on the SCET form factors. 

We adopt the calculational method of Refs.~\cite{Gritschacher:2013pha,Pietrulewicz:2014qza}, where the two loop graphs containing a ``secondary" massive quark bubble are calculated by starting with one-loop graphs describing the radiation of a massive gluon with mass $M$ and applying in a second step dispersion relations to account for the gluon splitting into a pair of secondary massive quarks with masses $m$.  The corresponding relation can be written as
\begin{align}
\label{eq:DispersionRelation}
\frac{(-i) g^{\mu \rho}}{p^2+i \epsilon} \Pi_{\rho \sigma}(m^2,p^2) \frac{(-i) g^{\sigma \nu}}{p^2 + i \epsilon} \; &= \; \frac{1}{\pi} \int \frac{\mathrm{d} M^2}{M^2} \frac{(-i) \bigg ( g^{\mu \nu} - \frac{p^{\mu} p^{\nu}}{p^2} \bigg)}{p^2 -M^2 + i \epsilon} \mathrm{Im} \left[ \Pi(m^2,M^2)\right] \nn \\
 &\quad
  - \frac{(-i) \bigg ( g^{\mu \nu} -\frac{ p^{\mu} p^{\nu}}{p^2} \bigg)}{p^2 + i\epsilon} \Pi (m^2 ,0) \, .
\end{align}
Here $\Pi(m^2,p^2)$ is the gluonic vacuum polarization due to the massive quark-antiquark bubble,
\begin{align}
\label{eq:Pidef}
\Pi^{AB}_{\mu \nu}(m^2,p^2) = -i (p^2 g_{\mu \nu} - p_{\mu} p_{\nu} ) \Pi(m^2,p^2) \delta^{AB} \equiv \int \mathrm{d}^4 x \, e^{i p x} \langle 0 | T J_{\mu}^A(x)J_{\nu}^B(0)| 0 \rangle \, ,
\end{align}
with the imaginary part in $d=4-2\epsilon$ dimensions given by
\begin{align}\label{eq:Im_Pi}
&\mathrm{Im}\!\left[\Pi(m^2,p^2)\right]
  = \theta(p^2\!-\! 4m^2) \, g^2 T_F  \left( \frac{p^2}{\tilde \mu^2} \right)^{\!-\epsilon} \,\frac{2^{3-2d}\pi^{(3-d)/2}}{\Gamma\Big(\frac{d+1}{2}\Big)}  
  \bigg(d\!-\! 2\!+\! \frac{4m^2}{p^2}\bigg)
  \!\bigg(1-\frac{4m^2}{p^2}\bigg)^{(d-3)/2} \! .
\end{align}
We note that the same method can be applied to account for any kind of secondary particles by a corresponding choice of the polarization function $\Pi$. 
Eq.~(\ref{eq:DispersionRelation}) allows us to express the contribution to the SCET form factor due to the massive quark loops as
\begin{align}
\label{eq:F2bare}
F_{\rm SCET}^{(C_F T_F , \, \mathrm{bare})}(Q, m, \Lambda) 
 &=  F_{\rm SCET}^{(\mathrm{OS},C_FT_F,\mathrm{bare})}(Q,m) 
  \, \nn \\
&
\quad 
- \bigg ( \Pi (m^2,0) - \frac{\, \alpha_s^{(n_l)}  (\mu) T_F}{3\pi}   
 \frac{1}{\epsilon}\bigg ) 
 F_{\rm SCET}^{(1, \, \mathrm{bare})} (Q, m, \Lambda) \, ,
\end{align}
where the ``on-shell'' form factor is
\begin{align}\label{eq:dispersive_OS}
 F_{\rm SCET}^{(\mathrm{OS},C_FT_F,\mathrm{bare})}(Q,m) 
  = \frac{1}{\pi}\,\int \frac{\df M^2}{M^2}\,F_{\rm SCET}^{(1,\, \mathrm{bare})}(Q,m,M) \,\mathrm{Im}\! \left[\Pi(m^2,M^2)\right] 
 \,.
\end{align}
In \eq{F2bare} $\Lambda$ denotes the gluon mass acting as our IR regulator, which we distinguish from the gluon mass $M$ used in the dispersion integration. Since total bare quantities are $\mu$-independent, we do not add $\mu$ as an argument to the components of bare quantities at a specific order. In $F_{\rm SCET}^{(\rm OS,\rm bare)}$ the massive quark contributions to the coupling are renormalized with the onshell subtraction, i.e. $F_{\rm SCET}^{(\mathrm{OS,bare})}$ is given in the scheme with $n_l$ dynamic flavors. In \eq{F2bare} the second term accounts for the change to $n_l+1$ dynamic flavors. The form factor itself is still unrenormalized, as indicated by the (bare) superscript. We perform the $\overline{\rm MS}$ renormalization for the SCET current using \eq{SCETamplitude}.
Incorporating Eqs.~(\ref{eq:F2bare}) and~(\ref{eq:SCETamplitude}) into Eq.~(\ref{eq:CmCFTF5f}) the result for $C_m^{(C_F T_F, \, n_l)}$ can be written as 
\begin{align} \label{eq:Cmstep1}
C_m^{(C_F T_F, \, n_l)}\Big(m,\frac{Q}{m},\mu\Big) 
& =F_{\rm SCET}^{(\mathrm{OS},C_FT_F,\mathrm{bare})}(Q,m)  
 \\
&\quad
 - \bigg ( \Pi (m^2,0) - \frac{\, \alpha_s^{(n_l)} (\mu) T_F}{3\pi} \frac{1}{\epsilon}\bigg )  \Big( F_{\rm SCET}^{(1)}(Q,m,\Lambda,\mu) -  Z_{\rm SCET}^{(1)}(Q,m,\mu) \Big)  \nn \\
& \quad 
 +  Z_{\rm SCET}^{(C_F T_F)}(Q,m,\mu) 
 - \frac{ \alpha_s^{(n_l)}(\mu)T_F}{3\pi}\, \ln \bigg (\frac{m^2}{\mu^2} \bigg) \,  F_{\rm SCET}^{(1)}(Q,m,\Lambda,\mu) \, . \nn
\end{align}
Here the 1-loop form factor $F_{\rm SCET}^{(1,\mathrm{bare})}$ is a UV and IR divergent amplitude, and $Z_{\rm SCET}^{(C_FT_F)}$ is the SCET current counterterm in the $(n_l+1)$-flavor scheme. Using the explicit form of $\Pi (m^2,0)$ in \eq{pi0} one can rewrite \eq{Cmstep1} as
\begin{align}
\label{eq:Cmfinal5f}
C_m^{(C_F T_F, \, n_l)}\Big(m,\frac{Q}{m},\mu\Big) 
 &=F_{\rm SCET}^{(\mathrm{OS},C_FT_F,\mathrm{bare})}(Q,m) +  Z_{\rm SCET}^{(C_F T_F)}(Q,m,\mu) 
  \\
 &\quad
   + \bigg ( \Pi (m^2,0) - \frac{\, \alpha_s^{(n_l)} T_F}{3\pi} \frac{1}{\epsilon}\bigg ) Z_{\rm SCET}^{(1)}(Q,m,\mu) 
  \,, \nn
\end{align}
where we see explicitly that the dependence on the IR regulator is canceled. Note that we could have also carried out the computation employing the $(n_l+1)$-flavor scheme to determine $C_m^{(C_FT_F,n_l+1)}$, which involves converting the bHQET form factor from the $n_l$ to $(n_l+1)$-flavor scheme. In this case the cancellation of IR divergences occurs in a different manner, and involves the ${\cal O}(\alpha_s)$ bHQET form factor. This approach is discussed in \app{Cmnlplus1}.

Note that nothing in \eq{Cmfinal5f} depends on the low energy bHQET theory. Therefore the result applies equally well to the case where one integrates out the heavy quark loop without approaching the jet invariant mass threshold $s_t\to m^2$ and matches onto a $n_l$-flavor SCET theory instead of bHQET. In this case the matching coefficient only contains the contribution from the massive quark loop and receives corrections starting at ${\cal O}(\alpha_s^2 C_F T_F)$, so switching between the $n_l$ and $(n_l+1)$-flavor schemes only affects the corrections at ${\cal O}(\alpha_s^3)$ and beyond. This is in close analogy to the case of primary massless quarks discussed in detail in Refs.~\cite{Gritschacher:2013pha,Pietrulewicz:2014qza}.

\subsection{One-loop computation for secondary massive gluons}
\label{sec:oneloopTFCF}

\begin{figure}
 \centering
 \includegraphics[width=0.8\linewidth]{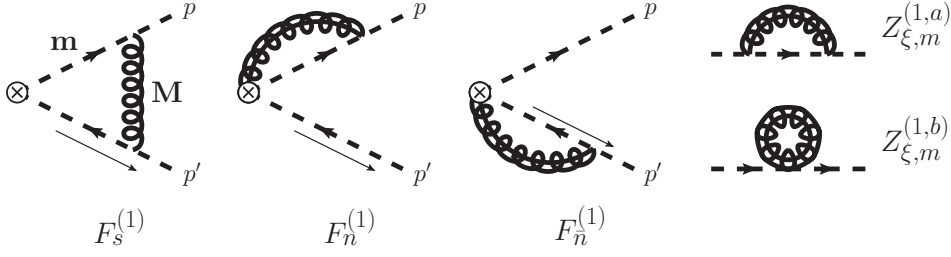}
 \caption{Non-vanishing EFT diagrams for the computation of the hard current at $\mathcal{O}(\alpha_s)$ with primary massive quarks and secondary massive gluons with masses $m$ and $M$, respectively. Soft-bin subtractions are implied for the collinear diagrams. 
 \label{fig:hardfunction_diagrams_m}
 }  
\end{figure}

Having laid out the basic framework in the previous section we now start with calculating the one loop SCET heavy quark form factors for a top-quark of mass $m$ with a massive gluon of mass $M$ to be used in the dispersion relation. The complete unrenormalized SCET result for the current form factor at $\mathcal{O}(\alpha_s)$ can be written as 
\begin{align}\label{eq:F_decompose}
F_{\rm SCET}^{(1,{\rm bare})}(Q,m,M)= F^{(1,{\rm bare})}_{\SCET,m=0}(Q,M) + \underbrace{F_{\rm SCET}^{(1,{\rm bare})}(Q,m,M)-F^{(1,{\rm bare})}_{\SCET,m=0}(Q,M)}_{=\,\delta F_{\SCET}^{(1,{\rm bare})}(m,M)} \,.
\end{align}
The correction with primary massless quarks $F^{(1,{\rm bare})}_{\rm SCET,m=0}$ has been already calculated in Refs.~\cite{Chiu:2007yn,Chiu:2007dg,Chiu:2009yx,Chiu:2012ir,Gritschacher:2013pha} and reads in $d = 4- 2\epsilon$ dimensions 
\begin{align}
\label{eq:FSCET_bare_d2}
F^{(1,{\rm bare})}_{\SCET,m=0} = \frac{\alpha_s(\mu) C_F}{4\pi} \bigg \{ \frac{2}{\epsilon^2} +\frac{3}{\epsilon}-\frac{2}{\epsilon} \lqp + (2 \lqp -3) L_M - L_M^2 +\frac{9}{2}-\frac{5\pi^2}{6} +\mathcal{O}(\epsilon) \bigg \} \, ,
\end{align}
where $\lqp = \ln \, (\frac{-Q^2 - i0}{\mu^2})$ and $L_M = \ln \, (\frac{M^2}{\mu^2})$. The corresponding one-loop counterterm in $\overline{\rm MS}$ reads
\begin{align}
\label{eq:ZcF1}
  Z_{\rm SCET}^{(1)} 
   = \frac{\alpha_s(\mu) C_F}{4\pi} \bigg \{\! -\frac{2}{\epsilon^2} -\frac{3}{\epsilon}+\frac{2}{\epsilon} \lqp \bigg \}
   \, .
\end{align}
Fig.~\ref{fig:hardfunction_diagrams_m} illustrates the SCET graphs with massive gluons needed to compute $F_{\rm SCET}^{(1,\mathrm{bare})}$. For the first three graphs in Fig.~\ref{fig:hardfunction_diagrams_m} the form factor contributions are defined as prefactors to the spinors, $F_{i}^{(1)} \,\bar{u}_{n,p}\gamma^\mu u_{\bar{n},p^\prime}$ for $i = n, \bar{n},s$ and are computed using the SCET Feynman rules for massive quarks given in Ref.~\cite{Leibovich:2003jd}. 

Due to the eikonal structure the result for the soft diagram, $F_{s}^{(1,\mathrm{bare})}$, is same as that for primary massless quarks [here $\tilde \mu^2 = \mu^2 e^{\gamma_E}/(4\pi)$],
\begin{align}
F_{s}^{(1,\mathrm{bare})} = & \,-2i g^2 C_F \tilde{\mu}^{2\epsilon} \int{\frac{\df^dk}{(2\pi)^d}\,\frac{1}{[k^-+i\epsilon]}\,\frac{1}{[k^+-i\epsilon]}}\,\frac{1}{[k^2-M^2+i\epsilon]} \,  .
\label{eq:Vs1b}
\end{align}
For the $n$-collinear diagram we get
\begin{align}
F_{n}^{(1,\mathrm{bare})}&=2i g^2 C_F\tilde{\mu}^{2\epsilon} \int \frac{\df^dk}{(2\pi)^d} \, \frac{Q-k^-}{[k^2-Qk^+-\frac{m^2}{Q}k^-+i\epsilon]}\,\frac{1}{[k^-+i\epsilon]}\,\frac{1}{[k^2-M^2+i\epsilon]} \, .
\label{eq:Vnm}
\end{align}
We can decompose this contribution into a correction corresponding to the diagram with primary massless quarks, and a UV and IR-finite difference of terms which can be computed in 4 dimensions,
\begin{align}\label{eq:Vnmdec}
F_{n}^{(1,\mathrm{bare})}= F_{n,m=0}^{(1,\mathrm{bare})}+ 
  \big(F_{n}^{(1,\mathrm{bare})}-F_{n,m=0}^{(1,\mathrm{bare})} \big) \, .
\end{align}
After performing a contour integration in $k^+$, carrying out the $k_\perp$-integration and rescaling the label momentum as $k^- \equiv  z Q$, the finite correction due to the mass of the primary quark yields 
\begin{align}\label{eq:dVnm}
& F_{n}^{(1,\mathrm{bare})}-F_{n,m=0}^{(1,\mathrm{bare})}  \\
& = - \frac{\alpha_s C_F}{2\pi}\,\Gamma\left(2-\frac{d}{2}\right)\left(\frac{\mu^2 e^{\gamma_E}}{M^2}\right)^{2-\frac{d}{2}}\int_0^1 \df z\, \frac{1-z}{z}\left[\left(1-z+\frac{m^2}{M^2} z^2\right)^{\frac{d}{2}-2}-\left(1-z\right)^{\frac{d}{2}-2}\right] \nn \\
& = \, \frac{\alpha_s C_F}{2\pi}\left[\ln\left(\frac{1+a}{2}\right)\ln\left(\frac{1-a}{2}\right)+\frac{1+a}{1-a}\,\ln\left(\frac{1+a}{2}\right)+\frac{1-a}{1+a}\,\ln\left(\frac{1-a}{2}\right)+1+\mathcal{O}(\epsilon)\right] 
 \, , \nn
\end{align}
with 
\begin{align}  \label{eq:a}
 a=\sqrt{1-\frac{4m^2}{M^2}} \, .
\end{align}
In SCET loop graphs include soft 0-bin subtractions~\cite{Manohar:2006nz} which ensure that there is no double counting of infrared regions. For the soft 0-bin subtraction of $F_{n}^{(1, \bare)}$ the dependence on the primary quark mass drops out, and we obtain the same result as for primary massless quarks, which is therefore fully contained in $F_{n,m=0}^{(1, \bare)}$. Note that the result in Eq.~(\ref{eq:dVnm}) does not contain any rapidity divergences, so that rapidity logarithms arise only in the computation of $F_{n,m=0}^{(1)}$.  This can be understood from the fact that the corrections due to soft modes are the same for massless and massive primary quarks, so that the rapidity divergences in the soft sector and, by consistency, also in the collinear sectors have to agree in both cases.

The $\bar{n}$-collinear diagram corresponds to switching $k^{-}$ and $k^{+}$ in Eq.~(\ref{eq:Vnm}). We perform a decomposition analogous to Eq.~(\ref{eq:Vnmdec}),
\begin{align}
F_{\bar{n}}^{(1,\mathrm{bare})}= F_{\bar{n},m=0}^{(1,\mathrm{bare})}+  \big(F_{\bar{n}}^{(1,\mathrm{bare})}-F_{\bar{n},m=0}^{(1,\mathrm{bare})} \big) \,.
\end{align}
The difference correction due to the primary quark mass is again UV and IR-finite and does not contain any rapidity divergences. Thus it yields for any choice of regulator the same result as the $n$-collinear correction, i.e.
\begin{align}
F_{\bar{n}}^{(1,\mathrm{bare})}-F_{\bar{n},m=0}^{(1,\mathrm{bare})}  =  F_{n}^{(1,\mathrm{bare})}-F_{n,m=0}^{(1,\mathrm{bare})} \, .
\end{align}
Finally, we also have to consider the wave function corrections. In analogy to the computation in Ref.~\cite{Fleming:2007xt} we have 
\begin{align}
 \Sigma^{(1)} =2ig^2  C_F\tilde{\mu}^{2\epsilon}\frac{\slashed{n}}{2}\int{\frac{\df^dk}{(2\pi)^d}\frac{Qm^2(3-\epsilon)-(Q^2k^+ +Q p^2+m^2 k^-)(1-\epsilon)}{Q^2[k^2-M^2+i\epsilon][(k+p)^2-m^2+i\epsilon]}} \, .
\end{align}
Using $p^2=m^2+\Delta^2$ and decomposing the integrals into elementary one- and two-point functions we obtain 
\begin{align}\label{eq:ZZ}
 \Sigma^{(1)}   =& \, i g^2  C_F\tilde{\mu}^{2\epsilon} \frac{\slashed{n}}{2} \, \frac{(1-\epsilon)}{Q(m^2+\Delta^2)}\bigg\{\left[A_0(m^2)-A_0(M^2)\right][2m^2+\Delta^2]\nn \\
&  + \, B_0(m^2+\Delta^2,M^2,m^2)\left[\frac{4m^2(m^2+\Delta^2)}{1-\epsilon}+2m^2M^2 +M^2\Delta^2-\Delta^4\right] \bigg\} \,,
\end{align}
which uses the loop integrals 
\begin{align}
  A_0(m^2) &= \int \frac{\df^dk}{(2\pi)^d} \, \frac{1}{[k^2-m^2+i\epsilon]} \,, \nn \\
 B_0(p^2,M^2,m^2) &=\int \frac{\df^dk}{(2\pi)^d} \, \frac{1}{[k^2-M^2+i\epsilon]}\, \frac{1}{[(p-k)^2-m^2+i\epsilon]} \, .
\end{align}  
The wave function renormalization constant $Z_{\xi}^{(1)}$ is defined by taking the on-shell limit $\Delta\to 0$ 
\begin{align}
 \Sigma^{(1)} \ \stackrel{\Delta\rightarrow 0}{\longrightarrow} \
  i \frac{\slashed{n}}{2} \, \frac{1}{Q} \left[2 m \, \delta m_M^{(\rm OS,1)} + \Delta^2  \, Z_{\xi}^{(1)}+\mathcal{O}(\Delta^4)\right] \, ,
\end{align}
where $\delta m_M^{(\rm OS,1)}$ is the one-loop renormalization constant for the quark mass $m$ in the pole mass scheme for the interaction with a massive gluon (with mass $M$). The wavefunction correction $Z_{\xi}^{(1)}$ can be written in terms of the wavefunction correction for primary massless quarks and a UV and IR finite remainder,
\begin{align}
 Z_{\xi}^{(1)}= Z_{\xi,m=0}^{(1)} + \big( Z_{\xi}^{(1)}-Z_{\xi,m=0}^{(1)} \big) \, .
\end{align}
The remainder contribution in $d=4$ dimensions reads
\begin{align}\label{eq:dZm}
Z_{\xi}^{(1)}-Z_{\xi,m=0}^{(1)}=& \, \frac{\alpha_s C_F}{4\pi} \, \frac{3}{2a(1-a^2)^2}\bigg[2(1+a)^4(2-a)\,\ln\left(\frac{1+a}{2}\right)
\nn \\
& -2(1-a)^4(2+a)\,\ln\left(\frac{1-a}{2}\right) 
 +  a\left(11-14a^2+3a^4\right)+\mathcal{O}(\epsilon)\bigg]\,,
\end{align}
where $a$ was given above in \eq{a}.

The complete finite correction at one-loop, which accounts for the mass of the primary quark is given by the sum of the terms from Eqs.~(\ref{eq:dVnm}) and (\ref{eq:dZm}), 
\begin{align}\label{eq:deltaF}
 \delta F_{\rm SCET}^{(1)}(m,M)
  &=2\left(F_{n}^{(1,\mathrm{bare})}-F_{n,m=0}^{(1,\mathrm{bare})}\right)(m,M)
  -\left(Z_{\xi}^{(1)}-Z_{\xi,m=0}^{(1)}\right)(m,M) \,.
\end{align}
This result will be used for our two-loop computation in the next section.

\subsection{Two-loop computation for secondary massive quarks}
 \label{sec:twoloopTFCF}

In this section we use the one-loop results from \sec{oneloopTFCF} to calculate the two-loop graph with the massive quark loop, and to determine the $C_F T_F$ contribution to $C_m$. First we compute $F_{\rm SCET}^{(\mathrm{OS}, C_FT_F, \mathrm{bare})}$ via \eq{dispersive_OS} using the one-loop result in Eq.~(\ref{eq:F_decompose}). Again we can decompose the two loop SCET form factor into a primary massless component and a correction for primary massive top quarks:
\begin{align}
F_{\rm SCET}^{(\mathrm{OS},C_FT_F,\mathrm{bare})}
  = F_{\SCET, \, m=0}^{(\mathrm{OS},C_FT_F,\mathrm{bare})}
    + \delta F_m^{(C_FT_F)}
\end{align}
The calculation for primary massless quarks has already been performed in Ref.~\cite{Pietrulewicz:2014qza}. We display the result here for convenience: 
\begin{align} \label{eq:FSCETOS}
F_{\SCET, \, m=0}^{(\mathrm{OS},C_FT_F,\mathrm{bare})}  &=
  \bigg ( \frac{\alpha_s^{(n_l)}(\mu)}{4 \pi} \bigg)^2 C_F T_F \,  \Bigg \{ \frac{2}{\epsilon^3}
  +\frac{1}{\epsilon^2} \bigg[ \frac{8}{3} L \minus 4\lqp 
     \plus \frac{8}{9} \bigg]
  + \frac{1}{\epsilon} \bigg[ \frac{4}{3} L^2 
  - \bigg( \frac{16}{3}L \plus \frac{16}{9} \bigg) \lqp 
  \nn \\ 
&\qquad
  \plus 4\lqp^2 \plus 4 L -\frac{65}{27}-\frac{\pi^2}{9}   \bigg]
 + \frac{56}{9} L^2 -\bigg[ \frac{242}{27}\plus \frac{4\pi^2}{9} \bigg] L -\frac{8}{3} \lqp^3 + \bigg[ \frac{16}{3}L \plus \frac{16}{9} \bigg] \lqp^2 
  \nn \\
&\qquad 
 - \bigg[ \frac{8}{3} L^2 + 8 L - \frac{130}{27}- \frac{2 \pi^2}{9} \bigg] \lqp + \frac{875}{54} + \frac{8 \pi^2}{9}-\frac{20 \zeta_3}{3}     \Bigg \}  
 \,.
\end{align}
The contribution from the two-loop $\overline{\rm MS}$ counterterm is known from the massless quark case and reads
\begin{align} \label{eq:ZCFTF}
Z^{(C_F T_F)}_{\rm SCET} =  \bigg ( \frac{\alpha_s^{(n_l+1)}(\mu)}{4 \pi} \bigg)^2 C_F T_F \, \bigg \{\!-\frac{2}{\epsilon^3} +\frac{1}{\epsilon^2} \bigg[ \frac{4}{3} \,\lqp - \frac{8}{9} \bigg] + \frac{1}{\epsilon} \bigg[ -\frac{20}{9}  \lqp +\frac{65}{27}+\frac{\pi^2}{3} \bigg] \bigg \} \, ,
\end{align}
where $L$ and $\lqp$ are defined in Eq.~(\ref{eq:logs}). The $1/\epsilon^n$ divergences in \eqs{FSCETOS}{ZCFTF}  differ, and are reconciled only once we account for the additional scheme change correction in the last term of \eq{F2bare}. The $\delta F^{(C_FT_F)}_{m}$ term can be computed by inserting the one-loop massive gluon correction term of Eq.~(\ref{eq:deltaF}) into the dispersive integral (\ref{eq:dispersive_OS}) which can be performed in four dimensions. The result reads
\begin{align}\label{eq:dF}
\delta F^{(C_F T_F)}_{m}=\bigg ( \frac{\alpha_s^{(n_l)}(\mu)}{4 \pi} \bigg)^2 C_F T_F \, \left\{\frac{1241}{81} - \frac{56 \pi^2}{27} + \frac{16}{3}\zeta_3\right\} \, .
\end{align}
Thus the only modification in the massive quark loop contributions to the form factor for primary massive quarks with respect to primary massless quarks is a simple constant term. In particular no additional rapidity logarithm $\sim  \ln(Q^2/m^2)$ appears, which can be again traced back to the universality of the soft corrections for massless and massive primary quarks.

Assembling all the pieces above in Eq.~(\ref{eq:Cmfinal5f}) we get the following result for $C_m^{(C_F T_F, \, n_l)} $:
\begin{align}
\label{eq:CmcFtFfinal}
C_m^{(C_F T_F, \, n_l)}\Big(m,\frac{Q}{m} ,\mu\Big) &= \bigg ( \frac{\alpha_s^{(n_l)}(\mu)}{4 \pi} \bigg)^2 C_F T_F \, \bigg \{ \! -\frac{8}{9} \lmu^3 - \frac{2}{9} \lmu^2+\bigg ( \frac{130}{27}+ \frac{2 \pi^2}{3}  \bigg) \lmu 
 \\ 
&\quad -\bigg ( \frac{4}{3}  \lmu^2+ \frac{40}{9}  \lmu+\frac{112}{27}  \bigg) \ln \, \bigg ( \frac{-Q^2\minus i0}{m^2}\bigg)     +\frac{5107}{162} -\frac{41 \pi^2 }{27} - \frac{4 \zeta_3}{9}   \bigg \} \, , \nn
\end{align}
which matches exactly with the $C_F T_F$ result we obtained above in \eq{Cm5fren}. In the next section we decompose the SCET form factor result into soft and collinear pieces in order to find the terms needed for the rapidity RGE analysis.

\subsection{Two Loop Ingredients for the Rapidity Renormalization Group}
 \label{sec:twoloopTFCFrge}

In order to determine the ingredients needed for the rapidity renormalization group analysis, we now calculate the $\mathcal{O}(\alpha_s^2 C_F T_F)$ SCET form factor contributions for the individual collinear and soft sectors using dispersion relations. We will employ the symmetric $\eta$-regulator~\cite{Chiu:2011qc,Chiu:2012ir} to regulate the rapidity divergences in the individual sectors. This corresponds to modifying the Wilson lines in the respective sectors according to
\begin{align}
 & W_n \, :  \, \frac{1}{\bar{n} \cdot \mathcal{P}} 
 \rightarrow
\frac{w^2(\nu)\, \nu^{\eta}}{ (\bar{n} \cdot \mathcal{P})^{1+\eta}}
  \, , \quad  
 & S_n \, : &  \, \frac{1}{n \cdot \mathcal{P}} \rightarrow \frac{1}{n \cdot \mathcal{P}} \, \frac{w(\nu)\, \nu^{\eta/2}}{|\bar{n} \cdot \mathcal{P} \minus n \cdot \mathcal{P}|^{\eta/2}} 
 \, , 
\end{align}
and similarly for $W_{\bar{n}}$ and $S_{\bar{n}}$. Here $\mathcal{P}^\mu$ denotes the label momentum operator and $w(\nu)$ is a dimensionless book keeping coupling parameter satisfying 
\begin{align}
\nu \frac{\df}{\df\nu} w(\nu) &= -\frac{\eta}{2} \, w(\nu) \,,
 & \lim_{\eta\to 0}  w(\nu) &=1 \,.
\end{align} 
The one-loop form factor corrections for the radiation of a massive gluon have been already calculated in Ref.~\cite{Chiu:2012ir} for massless quarks. Including the modification due to the quark mass in Eq.~(\ref{eq:deltaF}) they read after expanding in $\eta$ 
\begin{align}\label{eq:F1_Mns}
F^{(1,\bare)}_{\mathrm{SCET},\, n} & = F^{(1,\bare)}_{\mathrm{SCET},\,\bar{n}} 
  \\
& = \frac{\alpha_s^{(n_l+1)}(\mu) w^2(\nu) C_F}{4\pi}\,\Gamma(\epsilon)e^{\gamma_E \epsilon}
 \bigg(\frac{\mu^2}{M^2}\bigg)^{\epsilon}\left\{\frac{2}{\eta}+\ln\left(\frac{\nu^2}{Q^2}\right)+2 \psi(2-\epsilon) + 2\gamma_E -\frac{1-\epsilon}{2-\epsilon}\right\}
 \nn\\
 &\qquad +  \frac{\delta F_{\rm SCET}^{(1)}(m,M)}{2} 
\, , \nn  \\
F^{(1,\bare)}_{\mathrm{SCET},\,s} & = \frac{\alpha_s^{(n_l+1)}(\mu) w^2(\nu) C_F}{4\pi}\,\Gamma(\epsilon)e^{\gamma_E \epsilon}\bigg(\frac{\mu^2}{M^2}\bigg)^{\epsilon}\left\{-\frac{4}{\eta}-2\,\ln\left(\frac{\nu^2}{-M^2 + i0}\right)-2\psi(\epsilon) - 2\gamma_E \right\} 
 \, \nn . 
\end{align}
In the collinear results we have included the wave function contributions $Z_{\xi_n}/2$ and $Z_{\xi_{\bar{n}}}/2$. The soft-bin subtractions in the collinear diagrams vanish for the $\eta$-regulator. 

In direct analogy to Eq.~(\ref{eq:F2bare}) the corresponding two-loop expressions for the individual soft and collinear sectors read
\begin{align}
F_{\mathrm{SCET}, \, i}^{( C_F T_F , \mathrm{bare})}(Q,m)
  &= \,\frac{1}{\pi}\int \!\frac{\df M^2}{M^2} \, F^{(1,\bare)}_{\mathrm{SCET}, \,i}(Q,m,M)\,\mathrm{Im} \left[\Pi(m^2,M^2)\right] \nn
  \\
&\hspace{55pt} - \left(\Pi(m^2,0)-\frac{\alpha_s T_F}{3\pi} \,\frac{1}{\epsilon}\right)\! F^{(1,\bare)}_{\mathrm{SCET}, \,i}(Q,m,\Lambda)  \, . 
\end{align}
for $i=n,\bar{n},s$. Note that for this relation to make sense also the one-loop form factor corrections with a massless gluon have to be decomposed according to \eq{Cmsplit}. To achieve this goal we use a gluon mass $\Lambda \ll m$ as an infrared regulator which allows us to use the results in Eq.~(\ref{eq:F1_Mns}).  As discussed in \sec{HmSetup}, we absorb all divergences of the form $1/\eta$, $\eta^0/\epsilon^n$ in the form factors into separate counterterms $Z^{(C_F T_F)}_{\mathrm{SCET}, \, i}$ for each sector, so that
\begin{align}\label{eq:FcountertermsSplit}
F^{(1)}_{\mathrm{SCET},\, i} 
  &= F^{(1, \, \bare)}_{\mathrm{SCET},\, i} + Z^{(1)}_{\mathrm{SCET}, \, i}  
  \,, \qquad \qquad 
F^{(C_F T_F)}_{\mathrm{SCET},\, i} 
  = F^{(C_F T_F, \, \bare)}_{\mathrm{SCET},\, i} + Z^{(C_F T_F)}_{\mathrm{SCET}, \, i}  
\,. 
\end{align}
The explicit results for the counterterms at one-loop are given by\footnote{Although the full $\epsilon$-dependence in the expression proportional to $1/\eta$ should be in principle kept unexpanded, this is only relevant to ensure that the coefficient of the $1/\eta$ pole is explicitly $\mu$-independent, which is also true order by order in its $\epsilon$ expansion. Therefore we show here only the terms up to $\mathcal{O}(\epsilon^0)$ which contain the information we need later for the anomalous dimensions.}  
\begin{align}
\label{eq:Znnbar}
 Z^{(1)}_{\mathrm{SCET}, \, n}(Q,m,\Lambda,\mu,\nu)
  &=Z^{(1)}_{\mathrm{SCET}, \, \bar{n}} (Q,m,\Lambda,\mu,\nu) 
  \,, \\*
 &=   \frac{\alpha^{(n_l+1)}_s(\mu) \,w^2(\nu)\,C_F }{4\pi} \left \{ \frac{1}{\eta} \left [- \frac{2}{\epsilon} + 2\, \ln \bigg (\frac{\Lambda^2}{\mu^2} \bigg)  \right ] +\frac{1}{\epsilon} \left [-\frac{3}{2} - \ln \bigg (\frac{\nu^2}{Q^2} \bigg )\right ]  \right \} 
 \nn\\
 Z^{(1)}_{\mathrm{SCET}, \, s}(Q,m,\Lambda,\mu,\nu)
 &= \frac{\alpha^{(n_l+1)}_s(\mu) \,w^2(\nu)\,C_F }{4\pi} \left \{ \frac{1}{\eta} \left [ \frac{4}{\epsilon} -4 \,\ln \bigg (\frac{\Lambda^2}{\mu^2} \bigg)  \right ] -\frac{2}{\epsilon^2}+\frac{2}{\epsilon} \, \ln \bigg (\frac{\nu^2}{-\mu^2+i0} \bigg )  \right \} 
 \,, \nn
 \end{align}
 while at two-loop they read
 \begin{align}
 \label{eq:Znnbar2}
& Z^{(C_F T_F)}_{\mathrm{SCET}, \, n}(Q,m,\Lambda,\mu,\nu)= Z^{(C_F T_F)}_{\mathrm{SCET}, \, \bar{n}} (Q,m,\Lambda,\mu,\nu) \nn \\
& = \, \frac{\big[\alpha^{(n_l+1)}_s(\mu)\big]^2\,w^2(\nu) \, C_F T_F}{16\pi^2}\left\{\frac{1}{\eta}\left[-\frac{4}{3\epsilon^2}+\frac{20}{9\epsilon}+\frac{8}{3}L_m\,  \ln \bigg (\frac{\Lambda^2}{\mu^2} \bigg ) -\frac{4}{3}L_m^2-\frac{40}{9}L_m-\frac{112}{27} +\mathcal{O}(\epsilon)\right]\right. \nn \\
 & \hspace{0.5cm}\left.+\,\frac{1}{\epsilon^2}\left[-\frac{2}{3}\,\ln\left(\frac{\nu^2}{Q^2}\right)-1\right]+\frac{1}{\epsilon}\left[ \frac{10}{9}\,\ln\left(\frac{\nu^2}{Q^2}\right)+\frac{1}{6}+\frac{2\pi^2}{9}\right]\right\} \, , 
\nn\\*
& Z^{(C_F T_F)}_{\mathrm{SCET}, \, s}(Q,m,\Lambda,\mu,\nu) \nn \\
& = \, \frac{\big[\alpha^{(n_l+1)}_s(\mu)\big]^2\,w^2(\nu) \,C_F T_F}{16\pi^2}\left\{\frac{1}{\eta}\left[\frac{8}{3\epsilon^2}-\frac{40}{9\epsilon}-\frac{16}{3}L_m \, \ln \bigg (\frac{\Lambda^2}{\mu^2} \bigg )+\frac{8}{3}L_m^2+\frac{80}{9}L_m+\frac{224}{27}+\mathcal{O}(\epsilon) \right]\right. \nn \\
& \hspace{0.5cm} \left.-\,\frac{2}{\epsilon^3}+\frac{1}{\epsilon^2}\left[\frac{4}{3}\,\ln\left(\frac{\nu^2}{-\mu^2+i0}\right)+\frac{10}{9}\right]+\frac{1}{\epsilon}\left[-\frac{20}{9}\,\ln\left(\frac{\nu^2}{-\mu^2+i0}\right)+ \frac{56}{27}-\frac{\pi^2}{9} \right] \right\} \, .
\end{align}
Note that the sum $Z^{(C_F T_F)}_{\mathrm{SCET}, \, n}+Z^{(C_F T_F)}_{\mathrm{SCET}, \, \bn}+Z^{(C_F T_F)}_{\mathrm{SCET}, \, s}$ reproduces the result for the SCET current counterterm $Z^{(C_F T_F)}_{\mathrm{SCET}}$ in \eq{ZCFTF}. These results for the individual collinear and soft counterterms provide the necessary ingredients for determining the rapidity RGE for the collinear and soft sectors below in \sec{RRG}.

\section{Rapidity Evolution and Numerical Results}
\label{sec:RRGandnum}

\subsection{Rapidity Renormalization Group Evolution}
\label{sec:RRG}

In our result for the matching coefficient between bHQET and SCET at $\mathcal{O}(\alpha_s^2)$, given above in \eq{Hm5f}, we encountered a large logarithm $\alpha_s^2 C_F T_F\, \ln(m^2/Q^2)$. We discussed the setup for the resummation of such logarithms above in Sec.~\ref{sec:HmSetup}. As shown in  \sec{direct} these rapidity logarithms are only related to contributions of the virtual massive quarks that appear in the gluon vacuum polarization, and hence are the same as in the threshold corrections for massless primary quarks in Ref.~\cite{Pietrulewicz:2014qza}. There it was anticipated that they can be resummed by exponentiation, as is common for these kinds of logarithms. For example, for the radiation of a massive gauge boson the rapidity renormalization group implies that this exponentiation occurs to all orders in perturbation theory~\cite{Chiu:2007yn,Chiu:2007dg,Chiu:2012ir,Gritschacher:2013pha}.  The difference in our case is that the rapidity logarithms start at two-loops, and hence involve the additional issue of one-loop induced corrections due to the scheme change in the coupling constant.  

Here we will show explicitly how to treat the rapidity logarithms at $\mathcal{O}(\alpha_s^2 C_F T_F)$ in a rapidity renormalization group framework, and subsequently demonstrate that they indeed exponentiate. We start from Eq.~(\ref{eq:gammaCm}). Up to ${\cal O}(\alpha_s^2)$ we only have a contribution from the $C_F T_F$ dependent terms,
\begin{align}
\label{eq:CmResum}
\gamma^{C_m}_{\nu, \, i}(m,\mu) 
&=  \nu \frac{\df}{\df \nu} \ln\, Z_{{\rm SCET},i} - 
   \nu \frac{\df}{\df \nu} \ln\, Z_{{\rm bHQET},i} 
\nn\\
&=  \nu \frac{\df}{\df \nu}  Z_{{\rm SCET},i}^{(C_F T_F)} 
 - \frac{ \alpha_s^{(n_l)} (\mu) T_F}{3\pi}\, \ln \bigg (\frac{m^2}{\mu^2} \bigg) \,  \, \nu \frac{\df}{\df \nu}  Z_{\mathrm{SCET}, \, i}^{( 1)} +\mathcal{O}(\alpha_s^3)\, ,
\end{align}
where the second term accounts for coupling conversion from the $(n_l+1)$-flavor to $n_l$-flavor scheme. As before, in the $n_l$-flavor scheme the bHQET graphs give no contribution.
The results from \sec{twoloopTFCFrge} can now be used to compute this $\nu$-anomalous dimension. Using Eq.~(\ref{eq:Znnbar}) we can calculate the one-loop correction, 
\begin{align}
\nu \frac{\df}{\df \nu}  \,  Z_{\mathrm{SCET}, \, n}^{( 1)} =  \nu \frac{\df}{\df \nu}  \,  Z_{\mathrm{SCET}, \, \bar{n}}^{( 1)} = -\frac{1}{2} \, \nu \frac{\df}{\df \nu}  \,  Z_{\mathrm{SCET}, \, s}^{( 1)} =- \frac{\alpha_s^{(n_l+1)}(\mu) \,C_F}{2\pi} \, \ln\left(\frac{\Lambda^2}{\mu^2}\right) \, ,
\end{align}
which exhibits dependence on the infrared gluon-mass regulator $\Lambda$. 
The two-loop term above can be calculated using Eq.~(\ref{eq:Znnbar2}) which 
gives
\begin{align}
\nu \frac{\df}{\df \nu}  \, Z_{\mathrm{SCET}, \, n}^{( C_F T_F )} &= \nu \frac{\df}{\df \nu}  \, Z_{\mathrm{SCET}, \, \bar{n}}^{( C_F T_F )}
 = -\frac{1}{2} \, \nu \frac{\df}{\df \nu}  \,  Z_{\mathrm{SCET}, \, s}^{( C_F T_F )}
 \nn \\
&
 =\frac{[\alpha_s^{(n_l+1)}(\mu)]^2 C_F T_F}{16\pi^2}\left\{-\frac{8}{3}L_m \, \ln\left(\frac{\Lambda^2}{\mu^2}\right)+\frac{4}{3} L_m^2+\frac{40}{9} L_m+\frac{112}{27}\right\} \,, 
\end{align}
where $L_m$ is defined in \eq{logs}.
Together these results determine the $\nu$-anomalous dimensions:
\begin{align}
\label{eq:NuAnomalousDim}
\gamma^{C_m, \, C_F T_F }_{\nu, \, n}(m,\mu) 
 &= \gamma^{C_m, \, C_F T_F }_{\nu, \, \bar{n}}(m,\mu)
  =-\frac{1}{2} \, \gamma^{C_m, \, C_F T_F }_{\nu, \, s}(m,\mu)
 \nn\\
  &= \frac{[\alpha_s(\mu)]^2 C_F T_F}{16\pi^2}\left\{\frac{4}{3} L_m^2+\frac{40}{9} L_m+\frac{112}{27}\right\} \, .
\end{align}
Note that the IR regulator has canceled out, and that here the coupling $[\alpha_s(\mu)]^2$ can be taken in either the $n_l$ or $(n_l+1)$-flavor scheme since the anomalous dimension starts at ${\cal O}(\alpha_s^2)$ and the difference is higher order.
This result suffices for solving the $\nu$-RGE equations at NNLL order. Using Eq.~(\ref{eq:Hmfactorized}) and Eq.~(\ref{eq:GammaNus}) we can write an analog of Eq.~(\ref{eq:Hevol}) for the $\nu$-evolution of $H_m$. 
From \eq{Hmfactorized} we have
\begin{align} \label{eq:Hmfactorizedagain}
 H_{m}^{(n_l)}\bigg(m,\frac{Q}{m},\mu\bigg) 
  &= H_{m, \,n}^{(n_l)}\bigg(m,\mu,\frac{\nu}{Q}\bigg) \ 
  H_{m,\, \bar{n}}^{(n_l)}\bigg(m,\mu,\frac{\nu}{Q}\bigg)\
  H_{m,\, s}^{(n_l)}\bigg(m,\mu,\frac{\nu}{m}\bigg)   \,.
\end{align}
With rapidity evolution this becomes
\begin{align} \label{eq:HmRRG}
 & H_{m}^{(n_l)}\bigg(m,\frac{Q}{m},\mu; \nu_Q, \nu_m \bigg) 
 \\
& \qquad\qquad
  = H_{m, \,n}^{(n_l)}\bigg(m,\mu, \frac{\nu_Q}{Q}\bigg) \,
   H_{m,\, \bar{n}}^{(n_l)}\bigg(m,\mu,\frac{\nu_Q}{Q} \bigg)
V_{\rm RRG}(\nu_Q, \, \nu_m , \, \mu) \, H_{m,\, s}^{(n_l)}\bigg(m,\mu, \frac{\nu_m}{m} \bigg)  
  , \nn
\end{align}
where on the LHS the dependence on $\nu_Q$ and $\nu_m$ comes from truncating the resummed perturbation theory for objects on the RHS. 
Here the functions $H_{m,n}^{(n_l)} = H_{m,\bn}^{(n_l)}$ and $H_{m,s}^{(n_l)}$ are given up to ${\cal O}(\alpha_s^2)$ by
\begin{align}\label{eq:Hmn}
& H_{m,n}^{(n_l)}\Big(m,\mu,\frac{\nu_Q}{Q}\Big)
 =  1+\frac{\alpha_s^{(n_l)}(\mu) C_F }{4 \pi} \, \bigg (  \lmu^2 -\lmu + 4 +\frac{\pi^2}{6} \bigg ) \nn \\
& \qquad
+\bigg ( \frac{\alpha_s^{(n_l)}(\mu)}{4 \pi} \bigg)^2 C_F^2 \, \bigg \{  \frac{1}{2}\lmu^4-\lmu^3+ \bigg( \frac{9}{2}+\frac{\pi^2}{6} \bigg) \lmu^2- \bigg ( \frac{11}{2} -\frac{11 \pi^2}{6} + 24 \zeta_3 \bigg) \lmu  \nn \\
&\qquad\qquad
 +\frac{241}{8}+ \frac{13 \pi^2}{3} - 8 \pi^2 \log 2 - 6 \zeta_3-\frac{163 \pi^4}{360}   \bigg \}  \nn \\
&\qquad
+\bigg ( \frac{\alpha_s^{(n_l)}(\mu)}{4 \pi} \bigg)^2 C_A C_F \, \bigg \{ -\frac{11}{9} \lmu^3 + \bigg ( \frac{167}{18}- \frac{\pi^2}{3} \bigg) \lmu^2  - \bigg ( \frac{1165}{54}+ \frac{28 \pi^2}{9} - 30 \zeta_3 \bigg) \lmu  \nn \\
&\qquad\qquad  
 + \frac{12877}{648} + \frac{323 \pi^2}{108} + 4 \pi^2 \log 2 + \frac{89 \zeta_3}{9} - \frac{47 \pi^4 }{180}    \bigg \} \nn \\
&\qquad
 +\bigg ( \frac{\alpha_s^{(n_l)}(\mu)}{4 \pi} \bigg)^2 C_F n_l T_F \, \bigg \{  \frac{4}{9}\lmu^3 -\frac{26}{9}\lmu^2  + \bigg ( \frac{154}{27} + \frac{8 \pi^2}{9} \bigg) \lmu   -\frac{1541}{162} - \frac{37 \pi^2}{27} -\frac{52\zeta_3}{9}  \bigg \} \nn \\
&\qquad
 +\bigg ( \frac{\alpha_s^{(n_l)}(\mu)}{4 \pi} \bigg)^2 C_F T_F \, \bigg \{ 2 \lmu^2+\bigg ( \frac{2}{3}+ \frac{8 \pi^2}{9}  \bigg) \lmu
 +\frac{3139}{162} -\frac{4 \pi^2 }{3} + \frac{8 \zeta_3}{3}
 \nn \\
&\qquad\qquad 
  + \bigg (  \frac{4}{3}  \lmu^2+ \frac{40}{9}  \lmu+\frac{112}{27} \bigg) \ln \, \bigg ( \frac{\nu_Q^2}{Q^2}\bigg)    
    \bigg \}  
\, , \\
& H_{m,s}^{(n_l)}\Big(m,\mu,\frac{\nu_m}{m}\Big)
  =1+\bigg ( \frac{\alpha_s^{(n_l)}(\mu)}{4 \pi} \bigg)^2 C_F T_F \, \bigg \{ \frac{8}{9} \lmu^3 + \frac{40}{9} \lmu^2+\bigg ( \frac{448}{27}- \frac{4 \pi^2}{9}  \bigg) \lmu
 \nn \\
&\qquad\qquad 
  +\frac{656}{27} -\frac{10 \pi^2 }{27} - \frac{56 \zeta_3}{9}
  - \bigg (  \frac{8}{3}  \lmu^2+ \frac{80}{9}  \lmu+\frac{224}{27} \bigg) \ln \, \bigg ( \frac{\nu_m^2}{\mu^2}\bigg)  
    \bigg \}  \, ,  \label{eq:Hms}
\end{align}
and contain no large logarithms for $\mu\simeq m$, and for $\nu_Q \simeq Q$  and $\nu_m \simeq m$, respectively. 
The evolution factor $V_{\rm RRG}$ sums the rapidity logs between $\nu_m$ and $\nu_Q$, and is defined as follows
\begin{align}
\label{eq:VRRG}
V_{\rm RRG}( \nu_f , \nu_i , \mu) = \exp \bigg\{ \int_{\ln \, \nu_i}^{\ln \, \nu_f } \df \, \ln\,  \nu\, \Big[ \gamma^{C_m}_{\nu, \, s} + (\gamma^{C_m}_{\nu, \, s})^* \, \Big] \bigg\} \, . 
\end{align}
The general result for $V_{\rm RRG}$, and the result at NNLL, will be given below.

Similarly to the $\nu$-anomalous dimensions, we can also determine individual $\mu$-anomalous dimensions for the collinear and soft sectors, $i=n,s,\bn$,
\begin{align}
\gamma^{C_m}_{\mu, \, i} 
   = \mu \frac{\df}{\df \mu} \ln \, Z_{\mathrm{SCET}, \, i} 
   - \mu \frac{\df}{\df \mu} \ln \, Z_{\mathrm{bHQET}, \, i} \, .
\end{align}
Repeating the steps below \eq{CmResum} we find 
\begin{align}
\label{eq:MuAnomalousDim}
\gamma^{C_m,C_F T_F }_{\mu, \, n}\Big(m,\mu,\frac{\nu}{Q}\Big)
 & =\frac{[\alpha_s^{(n_l)}(\mu)]^2 C_F T_F}{16\pi^2}\left\{-\bigg(\frac{8}{3}L_m+ \frac{40}{9}\bigg)\ln\left(\frac{\nu^2}{Q^2}\right)-4L_m-\frac{2}{3}-\frac{8\pi^2}{9}\right\} 
 \nn \\
 &=\gamma^{C_m  (C_F T_F) }_{\mu, \, \bar{n}}\Big(m,\mu,\frac{\nu}{Q}\Big)
  ,\nn \\
\gamma^{C_m, C_F T_F }_{\mu, \, s}\Big(m,\mu,\frac{\nu}{m}\Big)
 & =\frac{[\alpha_s^{(n_l)}(\mu)]^2 C_F T_F}{16\pi^2} \left\{\bigg(\frac{16}{3}L_m+\frac{80}{9}\bigg)\,\ln\left(\frac{\nu^2}{-\mu^2+i0}\right)- \frac{224}{27}+\frac{4\pi^2}{9} \right\} \,,
\end{align}
whose sum yields the same result for the ${\cal O}(\alpha_s^2 C_F T_F)$ $\mu$-anomalous dimension of $C_m^{(n_l)}$ as the difference of  \eqs{gammabHQET}{gammaSCET}, 
\begin{align}
& \gamma^{C_m, \, C_F T_F }_{\mu, \, n}\Big(m,\mu,\frac{\nu}{Q}\Big)
 +\gamma^{C_m, \, C_F T_F }_{\mu, \, \bar{n}}\Big(m,\mu,\frac{\nu}{Q}\Big)
+\gamma^{C_m, \, C_F T_F }_{\mu, \, s}\Big(m,\mu,\frac{\nu}{m}\Big) 
  \nn\\
& \qquad 
 =\frac{ \big[\alpha_s^{(n_l)}(\mu)\big]^2 C_F T_F}{16\pi^2}\left\{ \bigg(\frac{16}{3}L_m+\frac{80}{9}\bigg)L_Q-8L_m-\frac{260}{27}-\frac{4\pi^2}{3}\right\} \nn \\
& \qquad 
  = \left[\gamma^{(n_l+1)}_{\rm SCET} - \gamma^{(n_l)}_{\rm bHQET}\right]^{(C_F T_F)} = \gamma^{C_m,C_F T_F }_{\mu}(Q,m,\mu)
  \, ,
\end{align}
with $L_m$ and $L_Q$ defined in~\eq{logs}.   

Eqs.~(\ref{eq:Hevol}) and~(\ref{eq:HmRRG}) together include the evolution connected to $H_m$ in the 2-dimensional $\mu$-$\nu$ plane, including that from invariant mass scales $\mu_m$ to $\mu_Q$, that from invariant mass scales $\mu_m$ to $\mu_{\rm final}$, and that from rapidity scales $\nu_Q$ to $\nu_m$. 
As demonstrated in Ref.~\cite{Chiu:2012ir} the combined $\mu$-$\nu$ evolution can be performed along any path and the path independence implies the consistency equation: 
\begin{align}
\label{eq:munuConsistency}
\mu \frac{\df}{\df \mu} \gamma^{C_m}_{\nu, \, i} 
  = \bigg ( \frac{\partial}{\partial \mu} + \beta(g) \, \frac{\partial}{\partial g} \bigg )\gamma^{C_m}_{\nu, \, i} 
  =  \nu \frac{\df}{\df \nu} \gamma^{C_m}_{\mu, \, i} \, .
\end{align}
However, similar to the example of the massive Sudakov form factor considered in Ref.~\cite{Chiu:2012ir} we can see from Eq.~(\ref{eq:NuAnomalousDim}) that $\gamma^{C_m}_{\nu, \, s}$ contains potentially large logarithms $\ln ( \mu / m)$ for an arbitrary path in $\mu$-$\nu$-space. This is resolved by a prior resummation exploiting the fact that the derivatives in \eq{munuConsistency} are proportional to the cusp anomalous dimension. Since $C_m$ is a matching coefficient between a $(n_l+1)$-flavor and $n_l$-flavor theory, we can express Eq.~(\ref{eq:munuConsistency}) in terms of the difference between the cusp anomalous dimensions $\Gamma_{\rm cusp} [\alpha_s] $ in the $(n_l +1)$ and $n_l$-flavor schemes.  So for $\gamma_{\nu,s}^{C_m}$ we obtain 
\begin{align}\label{eq:munuConsistencyCusp}
\mu \frac{\df}{\df \mu} \gamma^{C_m}_{\nu, \, s}
 &= \nu \frac{\df}{\df \nu} \gamma^{C_m}_{\mu, \, s} 
 = -2 \, \Big(\Gamma_{\rm cusp} [ \alpha_s^{(n_l+1)}]  - \Gamma_{\rm cusp} [ \alpha_s^{(n_l)}] \Big)  \nn \\
& = \frac{\alpha_s^2 C_F T_F}{16 \pi^2}\bigg ( \frac{32}{3} L_m +\frac{160}{9} \bigg ) +\mathcal{O}(\alpha_s^3) \, ,
\end{align}
which can be checked using the explicit perturbative expression of $\Gamma_{\rm cusp}[\alpha_s]$ up to two loops,
\begin{align}
\label{eq:Cusp}
\Gamma_{\rm cusp}[ \alpha_s^{(n_f)}] = \frac{\alpha_s^{(n_f)}}{4 \pi} 4 C_F + \bigg(\frac{\alpha_s^{(n_f)}}{4 \pi} \bigg )^2 4 C_F \bigg [ \bigg ( \frac{67}{9} - \frac{\pi^2}{3} \bigg) C_A - \frac{20 n_f}{9} T_F \bigg ] +\mathcal{O}(\alpha_s^3)\,.
\end{align}
Integrating \eq{munuConsistencyCusp} in $\mu$ we obtain the resummed result for $\gamma^{C_m}_{\nu, \, s}$, 
\begin{align}
\gamma^{C_m }_{\nu, \, s}(m,\mu)
 &= -2 \; \int^{\ln \, \mu}_{\ln \, m} \df \,\ln\, \mu^\prime \,  \bigg (\Gamma_{\rm cusp} [ \alpha_s^{(n_l+1)}(\mu^\prime)] - \Gamma_{\rm cusp} [ \alpha_s^{(n_l)}(\mu^\prime) ]  \bigg ) 
 + \gamma^{C_m}_{\nu,s}(m,m) 
\nn \\
&=-\; \Big( {\omega}^{(n_l+1)}(\mu,m) - {\omega}^{(n_l)}(\mu,m)\Big)
  +\gamma^{C_m}_{\nu,s}(m,m)  
 \, .
\end{align}
Here the integration constant $\gamma^{C_m}_{\nu,s}(m,m)$ is the correction in the anomalous dimension $\gamma^{C_m}_{\nu,s}$ that does not multiply a logarithm $\ln(\mu^2/m^2)$. We are now in the position to write down a general expression for $V_{\rm RRG}$. Using Eq.~(\ref{eq:VRRG}) we find the all orders result
\begin{align}\label{eq:V_RRG}
V_{\rm RRG}( \nu_Q , \nu_m , \mu) = \exp \Bigg\{\Big[  \omega^{(n_l+1)}(\mu,m) -\omega^{(n_l)}(\mu,m) - \gamma^{C_m}_{\nu,s}(m,m)  \Big]  \ln \bigg ( \frac{\nu_m^2}{\nu_Q^2} \bigg ) \Bigg \} \, .
\end{align}
At NNLL order with the counting $\alpha_s(\mu) \ln(\nu_m/\nu_Q) \sim 1$, we can expand this exponential to the first non-trivial order.
At the order we are working 
\begin{align}
 \gamma^{C_m}_{\nu,s}(m,m)  
  = - \frac{\big[\alpha_s^{(n_l+1)}(m)\big]^2 C_F T_F}{16 \pi^2} \; \frac{224}{27}  +\mathcal{O}(\alpha_s^3)\, ,
\end{align} 
as can be seen from Eq.~(\ref{eq:NuAnomalousDim}), where we have for definiteness employed the $(n_l+1)$-flavor scheme. The evolution function ${\omega}$ at NNLL accuracy reads
\begin{align}
 {\omega}^{(n_f)}(\mu,\mu_0) 
  &= -\frac{\Gamma_0}{\beta_0} \bigg \{ \ln \, r +\!
 \bigg ( \frac{\Gamma_1}{\Gamma_0} - \frac{\beta_1}{\beta_0} \bigg) \frac{ \alpha_s^{(n_f)}(\mu_0)}{4 \pi} (r \!-\! 1) 
  \\
  &\qquad\qquad
  +\! \bigg( \frac{\Gamma_2}{\Gamma_0} - \frac{\beta_1 \Gamma_1}{\beta_0 \Gamma_0} - \frac{\beta_2}{\beta_0} + \frac{\beta_1^2}{\beta_0^2} \bigg) \frac{ \big[\alpha_s^{(n_f)}(\mu_0)\big]^2}{32 \pi^2} 
 (r^2 \!-\! 1) \bigg \} 
 ,\nn
\end{align}
where $r = \alpha_s^{(n_f)}(\mu)/\alpha_s^{(n_f)}(\mu_0)$ and the coefficients $\beta_i$ and $\Gamma_i$ are evaluated with $n_f$ flavors.

To extend the analysis to N$^3$LL resummation, one needs the result for the $\nu$-anomalous dimension $\gamma^{C_m}_{\nu,s}(m,m)$ at $\mathcal{O}(\alpha_s^3)$, which can be inferred from the coefficient of the rapidity logarithm appearing in a related DIS calculation~\cite{Ablinger:2014vwa} due to consistency (see Ref.~\cite{Hoang:DIS}).

\subsection{Numerical Results }
\label{sec:nums}

In this section we explore the impact of the two-loop correction to the hard function $H_m$ on the differential cross section and the corresponding improvement to the perturbative uncertainties. To do this we examine the evolved hard function ${H}_{\rm evol}(Q,m,\mu_{\rm final};\mu_Q,\mu_m,\nu_Q,\nu_m)$ from \eq{Hevol}. This function fully captures the multiplicative contributions for the differential cross section factorization theorem in \eq{dijetobservable}, including the matching at $\mu_Q\simeq Q$ in $H_Q^{(n_l+1)}$, the RG evolution from $\mu_Q$ down to $\mu_m\simeq m$ in $U_{H_Q}^{(n_l+1)}$, the matching at $\mu_m$ encoded in $H_m$, and through $U_v^{(n_l)}$ the RG evolution from $\mu_m$ down to a scale $\mu_{\rm final}$ where the soft and jet functions are evaluated.\footnote{The soft or jet functions also contain an additional evolution which is not purely multiplicative~\cite{Fleming:2007qr}.  This evolution affects the shape of the $\df\sigma/\df s_t \df s_{\bar t}$ distribution and was evaluated up to NNLL$^\prime$ order in Ref.~\cite{Jain:2008gb}.} Since the ingredient that has not been previously analyzed is $H_m$ we focus our numerical study on the impact of this function and the associated reduction in the resulting $\mu_m$ dependence.  For $H_m^{(n_l)}(m,Q/m,\mu_m; \nu_Q,\nu_m)$ we employ \eq{HmRRG}, which provides a decomposition of this function into collinear and soft components, $H_{m,i}^{(n_l)}$ with $i=n,\bn,s$, plus a kernel $V_{\rm RRG}$ which carries out the RG evolution in rapidity from $\nu_Q\simeq Q$ to $\nu_m\simeq m$.

We begin by converting the result for the collinear and soft components  $H_{m,i}^{(n_l)}$ in Eqs.~(\ref{eq:Hmn}) and~(\ref{eq:Hms}) from the pole-mass scheme to the $\MS$ mass scheme with $n_l+1$ dynamic flavors via
\begin{align}
m_{\rm pole} = \bar{m}^{(n_l+1)}(\mu) \, \bigg ( 1 - \frac{\alpha_s^{(n_l+1)}(\mu) C_F}{4\pi} \big (3 L_m-4\big ) \bigg) + {\cal{O}}(\alpha_s^2) \,.
\end{align}
The $\MS$ scheme is an appropriate renormalon-free short distance mass scheme to be employed in the hard function $H_m$. For consistency we also convert the results in \eqs{Hmn}{Hms} to the $(n_l+1)$-flavor scheme for the strong coupling. Together this yields up to $\mathcal{O}(\alpha_s^2)$
 \begin{align}\label{eq:HmMSbarn}
 & H_{\bar{m},n}^{(n_l+1)}\Big(\bar m,\mu,\frac{\nu_Q}{Q}\Big)
    = 1+\frac{\alpha_s^{(n_l+1)}(\mu) C_F }{4 \pi} \, \bigg (  \lmubar^2 -\lmubar + 4 +\frac{\pi^2}{6} \bigg ) \nn \\
 & \qquad
 +\bigg ( \frac{\alpha_s^{(n_l+1)}(\mu)}{4 \pi} \bigg)^2 C_F^2 \, \bigg \{  \frac{1}{2}\lmubar^4-\lmubar^3- \bigg( \frac{15}{2}-\frac{\pi^2}{6} \bigg) \lmubar^2+ \bigg ( \frac{33}{2} +\frac{11 \pi^2}{6} - 24 \zeta_3 \bigg) \lmubar  \nn \\
 &\qquad\qquad
  +\frac{177}{8}+ \frac{13 \pi^2}{3} - 8 \pi^2 \log 2 - 6 \zeta_3-\frac{163 \pi^4}{360}   \bigg \}  \nn \\
 &\qquad
 +\bigg ( \frac{\alpha_s^{(n_l+1)}(\mu)}{4 \pi} \bigg)^2 C_A C_F \, \bigg \{ -\frac{11}{9} \lmubar^3 + \bigg ( \frac{167}{18}- \frac{\pi^2}{3} \bigg) \lmubar^2  - \bigg ( \frac{1165}{54}+ \frac{28 \pi^2}{9} - 30 \zeta_3 \bigg) \lmubar  \nn \\
 &\qquad\qquad  
  + \frac{12877}{648} + \frac{323 \pi^2}{108} + 4 \pi^2 \log 2 + \frac{89 \zeta_3}{9} - \frac{47 \pi^4 }{180}    \bigg \} \nn \\
 &\qquad
  +\bigg ( \frac{\alpha_s^{(n_l+1)}(\mu)}{4 \pi} \bigg)^2 C_F n_l T_F \, \bigg \{  \frac{4}{9}\lmubar^3 -\frac{26}{9}\lmubar^2  + \bigg ( \frac{154}{27} + \frac{8 \pi^2}{9} \bigg) \lmubar   -\frac{1541}{162} - \frac{37 \pi^2}{27} -\frac{52\zeta_3}{9}  \bigg \} \nn \\
 &\qquad
  +\bigg ( \frac{\alpha_s^{(n_l+1)}(\mu)}{4 \pi} \bigg)^2 C_F T_F \, \bigg \{ \frac{4}{3} \lmubar^3+\frac{2}{3} \lmubar^2+\bigg ( 6+ \frac{10 \pi^2}{9}  \bigg) \lmubar
  +\frac{3139}{162} -\frac{4 \pi^2 }{3} + \frac{8 \zeta_3}{3}
  \nn \\
 &\qquad\qquad 
   + \bigg (  \frac{4}{3}  \lmubar^2+ \frac{40}{9}  \lmubar+\frac{112}{27} \bigg) \ln \, \bigg ( \frac{\nu_Q^2}{Q^2}\bigg)    
     \bigg \}  
  = H_{\bar{m},\bar n}^{(n_l+1)}\Big(\bar m,\mu,\frac{\nu_Q}{Q}\Big)
 \, , \\
 & H_{\bar{m},s}^{(n_l+1)}\Big(\bar{m},\mu,\frac{\nu_m}{m}\Big)
   =1+\bigg ( \frac{\alpha_s^{(n_l+1)}(\mu)}{4 \pi} \bigg)^2 C_F T_F \, \bigg \{ \frac{8}{9} \lmubar^3 + \frac{40}{9} \lmubar^2+\bigg ( \frac{448}{27}- \frac{4 \pi^2}{9}  \bigg) \lmubar
  \nn \\
 &\qquad\qquad 
   +\frac{656}{27} -\frac{10 \pi^2 }{27} - \frac{56 \zeta_3}{9}
   - \bigg (  \frac{8}{3}  \lmubar^2+ \frac{80}{9}  \lmubar+\frac{224}{27} \bigg) \ln \, \bigg ( \frac{\nu_m^2}{\mu^2}\bigg)  
     \bigg \}  \, ,  \label{eq:HmMSbars}
 \end{align}
where $L_{\bar m}=\ln (  \bar{m}^2/ \mu^2 )$ and $\bar m = \bar m^{(n_l+1)}(\mu)$ is the $\overline{\rm MS}$ mass for $n_l+1$ active flavors. For the bHQET evolution function $U_v^{(n_l)}$, when using the $\overline{\rm MS}$ mass scheme, we expand the pole mass appearing in the anomalous dimension in \eq{gammabHQET} in terms of $\bar{m}_t(\bar{m}_t)$
to obtain
\begin{align}  \label{eq:gammabHQETmbar}
\gamma_{\rm bHQET}\Big(\frac{Q}{\bar m},\mu\Big)
&
= \frac{\alpha_s^{(n_l)}(\mu)C_F}{4\pi}\big[\!-4 L +4\big] +\bigg ( \frac{\alpha_s^{(n_l)}(\mu)}{4 \pi} \bigg)^{\!2} \bigg\{
n_l C_F T_F\bigg[\frac{80}{9} \bar L-\frac{80}{9}\bigg]
\nn \\
&\ \ 
\quad+ C_F C_A\bigg[-\bigg (\frac{268}{9}- \frac{4\pi^2}{3}    \bigg ) \bar L+ \frac{196}{9} - \frac{4\pi^2}{3} + 8 \zeta_3\bigg] \bigg\} 
 \nn\\
 & + \frac{32\, \alpha_s^{(n_l)}(\mu)\alpha_s^{(n_l)}(\bar m) C_F^2}{(4\pi)^2} 
 +\mathcal{O}(\alpha_s^3)\,,
\end{align}
where $\bar L=\ln[(-Q^2-i0)/{\bar m}^2]$. For the $\nu$-anomalous dimensions the $\MS$ results are obtained by the simple replacement $m\to \bar m$, since they start at two-loops.  For our central results below we use $\mu_m=\nu_m=\bar m_t$ and $\mu_Q=\nu_Q=Q$. 

For our numerical analysis of $H_{\rm evol}$ we employ scale choices that are appropriate to the peak region of the differential cross section within bHQET. We fix $Q=\mu_Q=1\,{\rm TeV}$, which is a possible c.m.~energy for a future linear collider, and $\mu_{\rm final} =5 \GeV$ corresponding to the scale of the soft radiation. We do not vary these two scales here since their impact and associated uncertainties have been analyzed elsewhere~\cite{Fleming:2007xt}. They matter only for the overall normalization and thus cancel in the normalized spectrum. In addition we use the $\overline{\rm MS}$ mass $\bar m_t(\bar m_t)=163\,{\rm GeV}$ or pole mass $m_t=171.8\,{\rm GeV}$ using the two-loop conversion, and $\alpha_s^{(5)}(m_Z)=0.114$~\cite{Blumlein:2006be,Abbate:2010xh} and using two-loop conversion at $\mu=\bar m_t$ to obtain $\alpha_s^{(6)}(\mu)$.  For results with RG evolution that sums large logarithms  we use the so called primed counting, i.e. our results at NLL$^\prime$ and NNLL$^\prime$ include NLL and NNLL evolution kernels together with the hard function boundary conditions at ${\cal O}(\alpha_s)$ and ${\cal O}(\alpha_s^2)$, respectively.~\footnote{Going from NNLL$^\prime$ to an even higher order in the resummation,  N$^3$LL, does not affect any of the conclusions in this section, and therefore, for convenience, we carry out our numerical analysis at NNLL$^\prime$.} For the rapidity evolution we use the expression in \eq{V_RRG}, and the default rapidity scales $\nu_Q=Q$ and $\nu_m=m_t$, where $m_t$ is either the $\MS$ mass $\bar m_t(\bar m_t)$  or the pole mass.

\begin{figure}[t!]
 \includegraphics[width=0.52\columnwidth]{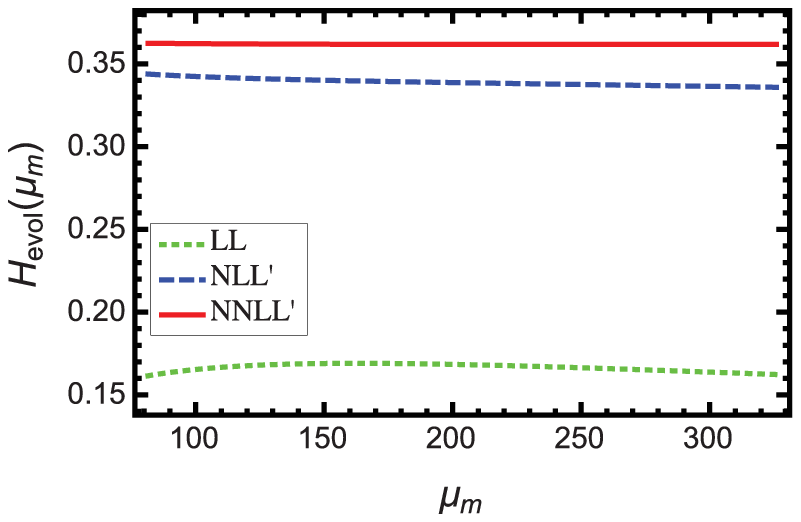}
\includegraphics[width=0.52\columnwidth]{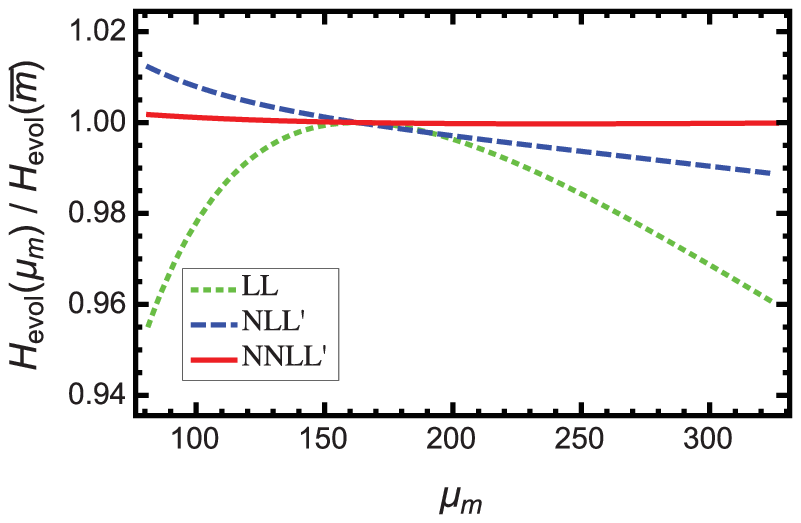} \\
\phantom{x}\hspace{-0.5cm} \includegraphics[width=0.52\columnwidth]{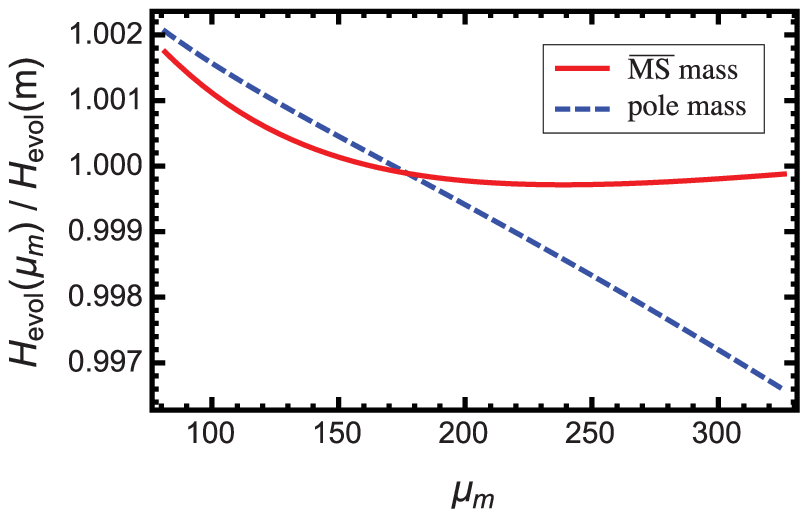} 
\hspace{-0.25cm} \includegraphics[width=0.52\columnwidth]{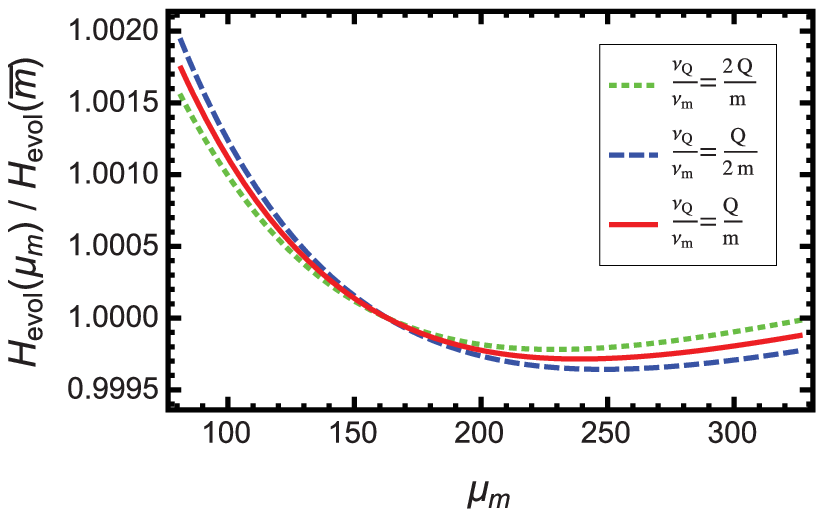} 
\caption{
Upper panels: Plots of the residual dependence on the matching scale $\mu_m$ for the unnormalized (left) and normalized (right) evolved hard function $H_{\rm evol}$ at three different orders in the evolution, using the $\MS$ mass. Lower left panel: Comparison of the scale dependence at NNLL$^\prime$ for the $\MS$ mass and the pole mass. Lower right panel: Impact of varying the ratio of rapidity scales $\nu_Q/\nu_m$ by a factor of two at NNLL$^\prime$ as a function of $\mu_m$, with the $\MS$ mass. }
\label{fig:munuvariation2}
\end{figure}

To determine the impact on the normalization we first note that the two-loop fixed order corrections to $H_{\bar m}^{(n_l+1)}$ turn out to be small, giving at the central scale $\mu_m=\bar m_t(\bar m_t)$ a 2\% correction and the fixed-order series
\begin{align} \label{eq:Hmnums}
  H_{\bar{m}}^{(n_l+1)}\Big(\bar m_t, \frac{Q}{\bar{m}_t},\mu_m=\bar m_t\Big) 
   =  1 +  0.126\text{(1-loop)} + 0.015\text{(2-loop)} = 1.141 \, .
\end{align} 
In the top-left panel of \fig{munuvariation2} we display the evolved hard function $H_{\rm evol}$ at the first three orders in resummed perturbation theory for values of $\mu_m$ in the range $\bar{m}_t/2<\mu_m<2 \bar{m}_t$. We use the $\overline{\rm MS}$ mass scheme and the expressions for $H_{\bar{m},n}^{(n_l+1)}$, $H_{\bar{m},\bar n}^{(n_l+1)}$ and $H_{\bar{m},s}^{(n_l+1)}$ from \eqs{HmMSbarn}{HmMSbars}. As already observed in Ref.~\cite{Fleming:2007xt}, there is a significant correction when going from LL to NLL$^\prime$ order which more than doubles $H_{\rm evol}$.  From NLL$^\prime$ to NNLL$^\prime$ we observe that the correction is notably smaller, indicating that the series has stabilized. Although the magnitude of these corrections is not captured by the $\mu_m$ variation, it is of the size expected from studying the uncertainty associated to the $\mu_{\rm final}$ variation. The complete study of the $\mu_{\rm final}$ variation requires including the jet and soft functions, which cancel the $\mu_{\rm final}$ dependence of $H_{\rm evol}$ to the order one is working.  We leave this for future work rather than taking it up here. We observe that the $\mu_m$ dependence significantly decreases as we go to higher order. This behavior is shown best in the top-right panel of \fig{munuvariation2}, where the same curves are plotted, but now normalized to $H_{\rm evol}(\mu_m=\bar{m}_t)$ at the respective order. The two-loop result for the hard function $H_{\bar m}^{(n_l+1)}$ plays a key role in this reduction of the scale dependence at NNLL$^\prime$. Note that the size of the $\mu_m$ variation of the blue dashed curve at $2\%$ correlates well with the size of the NNLO fixed order correction in \eq{Hmnums}, which gives a $+2\%$ correction.  Therefore it is reasonable to take the $\mu_m$ variation of the solid red curve in this figure as an estimate of the ${\cal O}(\alpha_s^3)$ correction in \eq{Hmnums}, which we take to be $\pm 0.2\%$.

In the lower-left panel of \fig{munuvariation2} we compare the dependence on $\mu_m$ at NNLL$^\prime$ for the $\MS$ mass with the corresponding result for the pole mass. In the pole mass case we employ \eqs{Hmn}{Hms} for $H_{m,n}^{(n_l)}$, $H_{m,\bar n}^{(n_l)}$ and $H_{m,s}^{(n_l)}$. We see that the pole mass exhibits a larger sensitivity to the renormalization scale $\mu_m$ implying a slightly slower convergence of the perturbative series, potentially related to IR renormalon effects. 

Finally, we can analyze the impact of the terms related to rapidity logarithms. For $\mu_m=\bar{m}_t(\bar{m}_t)$, these terms yield a numerical contribution of $-0.0014$ in the fixed-order full hard function $H_{\bar m}^{(n_l+1)}(\bar{m}_t, Q/\bar{m}_t, \mu_m=\bar{m}_t)$ in \eq{Hmnums}. Due to a relatively small coefficient, they do not give a significant correction in comparison with the remaining two-loop contributions which give a numerical correction of $0.0166$. Therefore, we anticipate the dependence on the rapidity scales $\nu_Q$ and $\nu_m$ to be rather mild.  In the lower-right panel of \fig{munuvariation2} we plot $H_{\rm evol}$ at NNLL$^\prime$ for the $\MS$ mass as a function of $\mu_m$, but now with three choices for $\nu_Q/\nu_m$. To obtain these results we varied $\nu_Q$ up and down by a factor of two, but we note that equivalent results are obtained by instead varying $\nu_m$ by a factor of two.  We see that varying $\nu_Q/\nu_m$ by a factor of 2 gives a negligible effect compared to the residual $\mu_m$ dependence at this order. Therefore, we conclude that including an uncertainty from $\nu$-variation is not necessary to obtain an estimate of the overall perturbative uncertainty of the cross section.

\section{Conclusions}
\label{sec:conclusions}

In the context of EFT factorization for boosted top quark production, we have extracted the hard function $H_m=|C_m|^2$ describing virtual fluctuations at the top-mass scale, completely at two-loop order using earlier results from Refs.~\cite{Bernreuther:2004ih, Gluza:2009yy}. This result provides the last missing ingredient needed to make N$^3$LL resummed predictions (up to the 4-loop cusp anomalous dimension) for the invariant mass distribution of top-jets in the peak region using the factorization theorem of Refs.~\cite{Fleming:2007qr,Fleming:2007xt} given in \eq{dijetobservable}. Particular focus was given to the contributions to $H_m$ from heavy quark loops, which induce terms with a large logarithm $\alpha_s^2 C_F T_F \ln(Q^2/m^2)$ that can not be treated with standard RG evolution in $\mu$. These terms were computed once more directly using collinear and soft matrix elements in SCET, and we have shown how they can be factorized using a rapidity cutoff $\nu$, and RG evolved using rapidity renormalization group equations. Interestingly, this factorization and RG evolution occurs within the Wilson coefficient $C_m$ and hence at the amplitude level.  Using our result for $H_m$ we have assessed the remaining perturbative uncertainty associated to the top-mass scale, $\mu_m\simeq m$, and estimate it to be very small, $\pm 0.2\%$, predicting that the two-loop result for $H_m$ provides a very accurate result for this function. The total normalization uncertainty in the differential cross section is expected to now be dominated by that from ${\cal O}(\alpha_s^3)$ perturbative corrections to the low-scale soft and jet functions, which could be estimated by a dedicated study of the residual $\mu_{\rm final}$ dependence at NNLL$^\prime$ order.

\acknowledgments{
This work was supported by the Office of Nuclear Physics of the U.S. Department of Energy (DOE) under Contract DE-SC0011090.  I.S. is also supported in part by the Simons Foundation Investigator grant 327942. The work of P.P. was supported by the German Science Foundation (DFG) under the Collaborative Research Center (SFB) 676 Particles, Strings and the Early Universe. We also thank the Erwin-Schrodinger Institute (ESI) for partial support in the framework of the ESI program “Jets and Quantum Fields for LHC and Future Colliders”.
}

\appendix

\section{Direct Calculation of $C_m$ in the $(n_l+1)$-flavor scheme} \label{app:Cmnlplus1}

In \sec{direct} we directly computed the ${\cal O}(\alpha_s^2 C_F T_F)$ massive quark correction to $C_m^{(n_l)}$ by using form factors in the $n_l$-flavor scheme. Since this coefficient lives at the border between the $(n_l+1)$ and $n_l$-flavor theories, we could just as well have carried out the calculation for $C_m$ by using form factors in the $(n_l+1)$-flavor scheme, and then converted to an $n_l$-flavor coupling at the very end. Of course the same result is obtained in this approach, but there are a few subtle differences in the calculation, which we discuss here.  

In particular, in \sec{direct} we noted that for the ${\cal O}(\alpha_s^2 C_F T_F)$ correction in the $n_l$-flavor scheme, the bHQET graphs  give no contribution. However, using the $(n_l+1)$-flavor scheme for the strong coupling this is no longer the case. To see this, consider the ratio in Eq.~(\ref{eq:Cmmethodtwo}) and express the denominator in the $(n_l +1)$-flavor scheme by inverting the decoupling relation given in Eq.~(\ref{eq:decouplingepsilon}):
\begin{align}
\label{eq:decoupling}
\alpha_s^{(n_l)}(\mu) = \alpha_s^{(n_l+1)}(\mu) \bigg [ 1+\frac{ \alpha_s^{(n_l+1)} (\mu) T_F}{3\pi}\, \ln \bigg (\frac{m^2}{\mu^2} \bigg) \bigg ] \, .
\end{align}
Expanding in $\alpha_s$ and using the notation in Eq.~(\ref{eq:ColorStructure}) we then get 
\begin{align}
\label{eq:CmCFTF}
C_m^{(C_F T_F, \, n_l+1)} \Big(m,\frac{Q}{m},\mu\Big)  &=\Big[ F_{\rm SCET}^{(C_F T_F , \,n_l+1)} (Q,m,\Lambda,\mu) 
  \\
& \qquad - \frac{ \alpha_s^{(n_l+1)} (\mu) T_F}{3\pi}\, \ln \bigg (\frac{m^2}{\mu^2} \bigg) \,  F_{\rm bHQET}^{(1,n_l)}\Big(\frac{Q}{m},\Lambda,\mu\Big) \Big]_{\alpha_s^{(n_l)} \rightarrow \alpha_s^{(n_l+1)}}
   \, . \nn
\end{align}
Here the second term comes from converting the strong coupling constant to $(n_l+1)$-flavors in the one-loop bHQET graph. Below we drop the flavors superscript on the form factors. Here it should be understood that all the terms are now expressed in the $(n_l+1)$-flavor scheme.
Then combining Eq.~(\ref{eq:CmCFTF}) and Eq.~(\ref{eq:F2bare}), and Eq.~(\ref{eq:SCETamplitude}) we get 
\begin{align}
\label{eq:Cmpreliminary}
C_m^{(C_F T_F, \, n_l+1)}\Big(m,\frac{Q}{m},\mu\Big) 
 &=  F_{\rm SCET}^{(\mathrm{OS},C_FT_F,\mathrm{bare})}(Q,m) 
  \\
 &\qquad 
 - \bigg ( \Pi(m^2,0) - \frac{\, \alpha_s^{(n_l+1)} (\mu) T_F}{3\pi} \frac{1}{\epsilon}\bigg ) F_{\rm SCET}^{(1, \mathrm{bare})}\Big(\frac{Q}{m},\Lambda\Big) 
   \nn \\
& \qquad
 + Z^{(C_F T_F)}_{\rm SCET}(Q,\mu) 
 - \frac{ \alpha_s^{(n_l+1)} (\mu) T_F}{3\pi}\, \ln \bigg (\frac{m^2}{\mu^2} \bigg) \,  F_{\rm bHQET}^{(1)}\Big(\frac{Q}{m},\Lambda,\mu\Big)
 \,. \nn
\end{align}
Note that both $F_{\rm SCET}^{(1, \mathrm{bare})}$ and $F_{\rm bHQET}^{(1)}$ are IR divergent. This result can be simplified by noting that in any flavor scheme the one-loop $C_m^{(1)}$ is given by the difference of one-loop renormalized SCET and bHQET amplitudes:
\begin{align}
\label{eq:Cm1ren}
C_m^{(1)}\Big(m,\frac{Q}{m},\mu\Big)
  &= F_{\rm SCET}^{(1)}(Q,m,\Lambda,\mu) - F_{\rm bHQET}^{(1)}\Big(\frac{Q}{m},\Lambda,\mu\Big) \, .
\end{align}
Using \eq{Cm1ren} in \eq{Cmpreliminary} we can then write
down a simpler expression for $C_m^{(C_F T_F,n_l+1)}$:  
\begin{align}
\label{eq:Cmfinal6f}
C_m^{(C_F T_F, \, n_l+1)} 
 &=F_{\rm SCET}^{(\mathrm{OS},C_FT_F,\mathrm{bare})}(Q,m)
  + Z^{(C_F T_F)}_{\rm SCET}(Q,\mu) 
  \nn \\
  &\qquad 
 - \bigg ( \Pi(m^2,0) - \frac{\, \alpha_s^{(n_l+1)} (\mu) T_F}{3\pi} \frac{1}{\epsilon}\bigg ) \Big(F_{\rm SCET}^{(1)}(Q,m,\Lambda,\mu) - Z^{(1)}_{\rm SCET}(Q,\mu) \Big)
 \nn \\
&\qquad
- \frac{ \alpha_s^{(n_l+1)} (\mu) T_F}{3\pi}\, \ln \bigg (\frac{m^2}{\mu^2} \bigg) \,  F_{\rm bHQET}^{(1)}\Big(\frac{Q}{m},\Lambda,\mu\Big)
  \nn \\
&=F_{\rm SCET}^{(\mathrm{OS},C_FT_F,\mathrm{bare})}(Q,m)
 + Z^{(C_F T_F)}_{\rm SCET}(Q,\mu) 
 \nn\\
&\qquad
 + \bigg ( \Pi (m^2,0) - \frac{\, \alpha_s^{(n_l+1)} (\mu) T_F}{3\pi} \frac{1}{\epsilon}\bigg ) Z_{\rm SCET}^{(1)}(Q,\mu)
 \nn \\
&\qquad
+\frac{ \alpha_s^{(n_l+1)} (\mu) T_F}{3\pi}\, \ln \bigg (\frac{m^2}{\mu^2} \bigg) \,  
C_m^{(1)}\Big(m,\frac{Q}{m},\mu\Big)
 \,.
\end{align}
This result can be used to compute $C_m^{(C_F T_F, \, n_l+1)} $. Comparing it with Eq.~(\ref{eq:Cmfinal5f}) we see that it can be rewritten as 
\begin{align} \label{eq:Cmconvert}
C_m^{(C_F T_F, \, n_l+1)} 
&=C_m^{(C_F T_F, \, n_l)} +\frac{ \alpha_s^{(n_l+1)} (\mu) T_F}{3\pi}\, \ln \bigg (\frac{m^2}{\mu^2} \bigg) \, C_m^{(1)} 
 \,,
\end{align}
and hence is fully consistent with determining $C_m^{(C_F T_F,n_l+1)}$ from  Eq.~(\ref{eq:Cmfinal5f}) and then simply applying the coupling conversion in Eq.~(\ref{eq:decoupling}) in the result. Note that in this $(n_l+1)$-flavor scheme approach the bHQET one-loop amplitude contributes and plays an important role in obtaining the scheme conversion term involving $C_m^{(1)}$ in the last line of \eq{Cmconvert}.

\section{bHQET current anomalous dimension at $\mathcal{O}(\alpha_s^3)$} 
\label{app:bHQET_anomdim}

To extent the resummation of large logarithms in the factorization theorem in \eq{dijetobservable} from NNLL to N$^3$LL the only missing ingredient -- besides the cusp anomalous dimension at four-loops --  is the  $\mathcal{O}(\alpha_s^3)$ noncusp anomalous dimension of the bHQET jet function or equivalently of the bHQET current (which are related to each other via \eq{consistency_bHQET} with the known three loop result for $\gamma_S$). The latter has not been so far given in the literature, but can be extracted from a recent result for the three-loop anomalous dimension of a cusped Wilson loop~\cite{Grozin:2014hna,Grozin:2015kna}, which is equivalent to the full anomalous dimension in HQET. Expanding their result in the lightlike limit $x \sim m/Q \to 0$, we obtain with the help of the Mathematica package HPL \cite{Maitre:2005uu}
\begin{align}  \label{eq:gammabHQET_3loop}
  \gamma_{\rm bHQET}\Big(\frac{Q}{m},\mu\Big)\bigg|_{\mathcal{O}(\alpha_s^3)}
    &
  =\bigg ( \frac{\alpha_s^{(n_l)}(\mu)}{4 \pi} \bigg)^{\!3} \bigg\{
 C_F C_A^2\bigg[\bigg (\!\!- \frac{490}{3} + \frac{536 \pi^2}{27} -\frac{88}{3}\zeta_3 - \frac{44 \pi^4}{45}   \bigg )L \\
 & \ + \frac{686}{9} - \frac{608 \pi^2}{27}  + \frac{1480}{9} \zeta_3 + \frac{44 \pi^4}{45}  + 
  \frac{8 \pi^2}{3} \zeta_3 - 72 \zeta_5
 \bigg] \nn \\
 & \ + C_F C_A T_F n_l \bigg[\bigg (\frac{1672}{27}-\frac{160\pi^2}{27} + \frac{224}{3} \zeta_3\bigg)L -\frac{712}{27} + \frac{160 \pi^2}{27} 
- \frac{992}{9} \zeta_3 \bigg] \nn \\
 & \ + C_F^2 T_F n_l \bigg[\bigg(\frac{220}{3}-64\zeta_3 \bigg)L -\frac{220}{3}+64\zeta_3\bigg]  + C_F (T_F n_l)^2\bigg[\frac{64}{27}L -\frac{64}{27}\bigg]\bigg\} \,, \nn
\end{align}
where $L=\ln[(-Q^2-i0)/m^2]$. The coefficient of this logarithm is proportional to the well-known lightlike cusp anomalous dimension at three loops, $\Gamma_{\rm cusp}^{(3)}$, while the non-logarithmic ingredient of \eq{gammabHQET_3loop} represents the noncusp part. Together with the corresponding anomalous dimension of the SCET current this enables one to predict the logarithmic structure of $H_m$ at three loops by solving \eq{muRGE}. Furthermore it allows one to extract the last missing ingredient to predict the full IR-divergent structure of the three-loop full QCD form factor for massive quarks for $m \ll Q$, which is for example in Ref.~\cite{Gluza:2009yy} the coefficient $K^{(3)}$ in Eq.~(63). 

\bibliographystyle{JHEP}
\bibliography{top3}

\end{document}